\documentclass[11pt]{article}

\usepackage{natbib}

 \usepackage{setspace} 
 \usepackage{array,booktabs}% http://ctan.org/pkg/{array,booktabs}

\newenvironment{sciabstract}{%
\begin{quote} \bf}
{\end{quote}}

\makeatletter
\def\fixedlabel#1#2{%
  \@bsphack%
  \protected@write\@auxout{}%
         {\string\newlabel{#1}{{#2}{\thepage}}}%
  \@esphack}

\newcounter{lastnote}

\usepackage{amsthm,amsmath}
\RequirePackage[OT1]{fontenc}
\RequirePackage{amsthm,amsmath}

\RequirePackage[colorlinks,citecolor=blue,urlcolor=blue]{hyperref}
\usepackage[latin1] {inputenc}
\usepackage{vmargin}
\usepackage{multirow}
\usepackage{subfigure}
\usepackage[normalem]{ulem}	
\usepackage[pagewise]{lineno}

\usepackage{caption}
\usepackage{fancyhdr}
\usepackage{amsthm, dsfont, tipa}
\usepackage{graphics}
\usepackage{graphicx,float,multirow}
\usepackage{tikz} 
\usetikzlibrary{arrows,decorations.markings}

 \tikzset{
  big arrow/.style={
    decoration={markings,mark=at position 1 with {\arrow[scale=2,#1]{>}}},
    postaction={decorate},
    shorten >=0.4pt},
  big arrow/.default=black}
  
    \tikzset{
  dash arrow/.style={
    decoration={markings,mark=at position 1 with {\arrow[scale=1.2,#1]{>}}},
    postaction={decorate},
    shorten >=0.4pt},
  dash arrow/.default=black}
  
  \tikzset{ doppia/.style={
  color=black, cylinder, draw, shape border rotate=90, aspect=0.15, text width=5.5em, minimum height=6.5em, minimum width=2em, align=center}}

\theoremstyle{plain}

\newtheorem{lemma}{Lemma}[section]

\usepackage{amssymb}
\usepackage{epsfig}
\usepackage[makeroom]{cancel}
\DeclareMathAlphabet{\mathpzc}{OT1}{pzc}{m}{it}
\newcommand{\LR}{\textrm{LR}}

\title{Bayesian approach to LR assessment in case of rare type match: careful derivation and limits}

\author
{Giulia Cereda,$^{1\ast, 2}$ \\
\\
\normalsize{$^{1}$University of Lausanne, Faculty of Law, Criminal Justice and Public Administration}\\
\normalsize{$^{2}$Leiden University, Mathematical Institute}\\
\\
\normalsize{$^\ast$To whom correspondence should be addressed; E-mail: giulia.cereda@unil.ch.}
}

\date{}

\begin{document} 

% Double-space the manuscript.

\baselineskip18pt

% Make the title.

\maketitle

% Place your abstract within the special {sciabstract} environment.
%%%%%%%%%%%%%%%%% ABSTRACT %%%%%%%%%%%%%%%%
\begin{sciabstract}
The likelihood ratio (LR) is largely used to evaluate the relative weight of forensic data regarding two hypotheses and for its assessment Bayesian methods are widespread in the forensic field.
However, the Bayesian `recipe' for the LR presented in most of literature consists in plugging-in Bayesian estimates of the involved nuisance parameters into a frequentist-defined LR: frequentist and Bayesian methods are thus mixed, giving rise to solutions obtained by hybrid reasoning. This paper provides the derivation of a proper Bayesian approach to assess LR for the `rare type match problem', the situation in which the expert wants to evaluate a match between the profile of a suspect and that of a trace from the crime scene, and this profile has never been observed before in the database of reference.
Bayesian LR assessment using the two most popular Bayesian models (beta-binomial and Dirichlet-multinomial) is discussed and compared to corresponding plug-in versions.
\end{sciabstract}

\section{Introduction}\label{Intro}

One of the main challenges of forensic science is that of properly evaluating the match between the characteristics of a crime stain (for instance a DNA profile) and the corresponding characteristics of some material from a known source (for instance from a suspect).
Typically, a couple of mutually exclusive hypotheses is defined, of the kind of `the crime stain came from the suspect' ($h_p$) and `the crime stain came from an unknown donor' ($h_d$). 

The forensic expert is given some data $D$ which can typically be split into \emph{evidence}, data directly related to the crime, and \emph{background}, additional data not directly related to the crime and only pertaining some nuisance parameter $\theta$ involved in the assessment of the likelihood ratio. 
%This is partially different from the `background information' $I$ as defined in \citet{aitken:2004, taroni:2014}, but we can say that often background data is part of the background information.
Evidence and background data will be modeled through random variables $E$ and $B$ respectively.
In particular, we are interested in the situation in which the forensic expert is asked to evaluate the match between a DNA profile of a suspect and the DNA profile of a stain found at the crime scene. It is intuitive to understand that (one of) the nuisance parameter(s) involved in this evaluation is the proportion of people with the same profile among the possible perpetrators: the more this profile is rare the more the suspect is in trouble. This parameter is unknown and thus the expert is given (or asks for) a database containing a list of DNA profiles from a sample from the population of possible perpetrators.
\textcolor{black}{The main difference between the frequentist and the Bayesian methodology is that the first considers the nuisance parameter $\theta$ and the correct hypothesis $h$ as fixed (without distribution) unknown quantities, while the second models the expert's uncertainty about their value through random variables, whose prior distributions reflect prior expert's beliefs.}

The largely accepted method for evaluating the data in order to discriminate between the two hypotheses of interest, is the calculation of the \emph{Bayes factor} (BF), regularly called in forensic context \emph{likelihood ratio} (LR) and defined as the ratio of the probability of observing the data under the two competing hypotheses:
\begin{equation*}
\label{LR}
\LR=\frac{\Pr(E=e,B=b \mid H= h_{p})}{\Pr(E=e,B=b \mid H=h_{d})}.
\end{equation*}

In the Bayesian framework (the one of interest for this paper) $\Pr$ is the joint distribution of all the random variables  in the model ($E$, $B$, $H$, and $\Theta$). 

On the other hand, frequentists, which consider $\theta$ and $h$ as  fixed quantities, use a different probability (here denoted as $\mathpzc{Pr}$) which can be expressed in terms of the Bayesian $\Pr$, in the following way: $\mathpzc{Pr}(\cdot):=\mathpzc{Pr}^\theta_{h}(\cdot)=\Pr(\cdot\mid \Theta=\theta, H=h)$. Thus, the frequentist likelihood ratio (denoted as $\mathpzc{LR}$) is defined as
$$
\mathpzc{LR}=\frac{\mathpzc{Pr}^\theta_{h_p}(E=e, B=b)}{\mathpzc{Pr}^\theta_{h_d}(E=e, B=b)}=\frac{\Pr(E=e, B=b\mid \Theta=\theta, H=h_p)}{\Pr(E=e, B=b\mid \Theta=\theta, H=h_d)}.
$$

Depending on the preferences of the expert, frequentist or Bayesian likelihood ratios can be used for the evaluation of forensic data. Once a choice has been made, it is important to be consistent with it, while literature often mixes up the two. To our knowledge, this paper and \citep{cereda:2015b} constitute the only forensic literature that discuss the differences between the two approaches.
 \citep{cereda:2015b} is concerned with the theoretical foundations of frequentist solutions, while this paper provides a simple and careful derivation of the proper Bayesian LR, for the rare type match problem (described in Section~\ref{rare}): the situation in which the DNA profile of the crime stain and that of the suspect match but they are not among the DNA profiles observed in the reference database.  
In Section~\ref{expo} we will discuss the fact that influential Bayesian forensic literature \textcolor{red}{\citep{weir:1996, aitken:2004, taroni:2010, taroni:2014, sjerps:2015}} seems to suggest the use of frequentist defined likelihood ratio ($\mathpzc{LR}$) and use Bayesian methodologies only inasmuch they provide a Bayesian estimate of $\theta$ to be plugged into $\mathpzc{LR}$. Others \citep{curran:2005, vanderhout:2015}, treat the likelihood ratio as function of $\theta$ and provide its posterior distribution with respect to the posterior distribution of $\theta$ given the data.  However, one of the main points of discussion is that there is no need of such hybrid derivations since the proper Bayesian LR is often very easy to obtain: this paper shows how this should be done, taking advantage of a very useful Lemma, presented in Section~\ref{lem}.
However, for this method to be advisable, the Bayesian prior should be chosen in a sensible way, reflecting the expert's opinion, and not by mathematical convenience as often happens.

%\textcolor{red}{to pass from the `estimation' of the LR via the plugging in of an estimate of $\theta$, to the downright  Bayesian `assignment' of the LR.} 
The two most common Bayesian models (beta-binomial and Dirichlet-multinomial) are discussed in Sections~\ref{beta} and \ref{diri}.
They are general enough to be applied to different kinds of forensic evidence evaluation, but will be here applied to DNA profiles obtained using the Y-STR marker system, with the double aim of exploring the performance of the conventional Bayesian prior choices for the rare type match case for non autosomal DNA, and of showing how a full Bayesian LR is to be defined and calculated.
Sensitivity analysis and comparison with proposed hybrid plug-in solutions are carried out.
%It is important to mention that the proposed models, the beta-binomial and the Dirichlet-multinomial, are the classical and most used Bayesian models in forensic statistics, due mostly to mathematical convenience (and/or convention). Their notoriety is one of the reason we decided to start to study and discuss a rigorous use of these two models. 
We are not entirely satisfied with the performance of classical models for the rare haplotype problem, which we believe would need different kinds of prior, more realistic and tuneable, such as those proposed in \citet{cereda:2015c}.
\subsection{Notation}
Throughout the paper the following notation is chosen: random variables and their values are denoted, respectively, with uppercase and lowercase characters: $x$ is a specific realisation of $X$. Random vectors and their values are denoted, respectively, by uppercase and lowercase bold characters: $\mathbf{p}$ is a realisation of the random vector $\mathbf{P}$. Bayesian probability is denoted with $\Pr(\cdot)$, while density of a continuous random variable $X$ is denoted by $p(x)$. For a discrete random variable $Y$, the continuous notation $p(y)$ and the discrete one $\Pr(Y=y)$ will be alternately used. Frequentist probability will be denoted as $\mathpzc{Pr}$.

%%%%%%%%%%%%%%%%% SECTION 1  %%%%%%%%%%%%%%%%

\section{The rare type match problem}\label{rare}
 
The DNA sequence of an individual is a very long sequence of letters (each corresponding to 4 nucleotides) which code the genetic instruction necessary for the life of the individual. The entire sequence is unique to each individual (with the only exception of homozygous twins, which share the same sequence), but a Y-STR DNA profile (also called \emph{haplotype}) usually consists on a short list of integers (typically 7 to 23) that describes only certain characteristics of the DNA sequence of the individual on the Y chromosome \citep{gill:2001}. Moreover, the Y-STR profile is shared between men in the same patrilineal lineage. For these reasons there is a positive probability that two different persons share the same profile. This is why we need to weight how probable is the observed match under the hypothesis that the suspect left the stain against how likely is the match under the hypothesis that someone else left the stain. 
Clearly, assuming that a match is always detected correctly (no false positives), the first probability is 1, and the second depends on the proportion $\theta$ of that profile among the population of potential perpetrator. 
Moreover, we are given a list of profiles from a sample of individuals belonging to the population of possible perpetrators to assess this frequency. Problems arise when the observed frequency of this characteristic is 0, the so-called `rare type match problem'.
This problem is particularly significant in case a new kind of forensic evidence for which the available database size is still limited is involved. This is the case, for instance, when using DIP-STR markers \citep{cereda:2014b}).
The same happens when Y-chromosome
(or mitochondrial) DNA profiles are used, since the set of possible haplotypes is extremely
large, and the coverage of available databases often limited.
%because of the lack of recombination involved when
%offspring DNA is generated from the DNA of the parents, each haplotype must be treated as a unit (the
%matching probability can't be obtained by multiplication across loci) so that
The case of Y-STR DNA will thus be retained here as an extreme but in practice common and important way in
which the problem of assessing the evidential value of rare type match can arise. 
This is a very appropriate and paradigmatic example, since literature provides examples of different approaches to evaluate the evidential value of rare Y-STR profile match \citep[][e.g.]{roewer:2000, andersen:2013b}, even though, in our opinion, a proper Bayesian derivation for the LR in rare type match case hasn't been proposed yet.
This problem is so substantial that it has been defined ``the fundamental problem of forensic mathematics'' by \citet{brenner:2010}.
We will now review some of the methods proposed by literature. Most of them address the problem of assessing the frequency of a type with zero occurrence, sometimes under the name of `zero numerator problem' \citep[e.g.][]{winkler:2002}. Notice that this is related, but not equivalent, to the problem of assessing the likelihood ratio in case of a rare type match. 
The \emph{empirical frequency estimator}, also called \emph{naive estimator}, that uses the frequency of the characteristic in the database, puts unit probability mass on the set of already observed characteristics, and it is thus unprepared for the observation of a new type. 
A solution could be the \emph{add-constant} estimators (in particular the well known \emph{add-one} estimator, due to \citet{laplace:1814}, and the \emph{add-half} estimator of \citet{krichevsky:1981}), which add a constant to the count of each type, included the unseen ones. However, this method requires to know the number of possible unseen types, and does not perform well when this number is large compared to the sample size (see \citet{gale:1994} for additional discussion). 
Moreover, \citet{Louis:1981} proposes the so-called `rule of three', that states that if $n$ is the size of the database, $3/n$ is a good approximation of the $95\%$ upper bound for the frequency. This is also proposed in a Bayesian framework, by \citet{jovanovic:1997, winkler:2002, chen:2008}.
Alternatively, \citet{good:1953}, based on an intuition on A.M. Turing, proposed the nonparametric \emph{Good Turing estimator} for the total unobserved probability mass, based on the proportion of singleton observations in the sample. An extension of this estimator is applied to the LR assessment in the rare type match in \citet{cereda:2015b}. 
For a comparison between \emph{add one} and \emph{Good-Turing} estimator, see \citet{orlitsky:2003}.
As pointed out in \citet{anevski:2013}, the \emph{naive estimator}, and the \emph{Good Turing estimator} are
in some sense complementary: the first gives a good estimate for the observed types, and the second for the probability mass of the unobserved ones. 
More recently, \citet{orlitsky:2004} have introduced the \emph{high profile estimator}, which extends the tail of the \emph{naive estimator} to the region of unobserved types. \citet{anevski:2013} improved this estimator and provided the consistency proof. 
Papers that address the rare Y-STR haplotype problem in forensic context are for instance \citet{egeland:2008}, \citet{brenner:2010}, and \citet{cereda:2015, cereda:2015c}. Moreover, the Discrete Laplace method presented in \citet{andersen:2013b}, even though not specifically designed for the rare type match can be successfully applied to that extent \citep{cereda:2015b}.
Bayesian nonparametric estimators for the probability of observing a new type have been proposed by \citet[e.g.][]{tiwari:1989, lijoi:2007, favaro:2009}.
However, for the likelihood ratio assessment it is required not only the probability of observing a new species but also the probability of observing this same species twice (according to the defense the crime stain profile and the suspect profile are two independent observations). \citet{cereda:2015c} is the first paper that addresses the problem of LR assessment in the rare haplotype case using Bayesian nonparametric models. 
%As prior we will use the two parameter Poisson Dirichlet distribution, which is proving useful in many discrete domain, in particular language modelling \citep{teh:2006}. 
%%%%%%%%%%%%%%%%% SECTION 2  %%%%%%%%%%%%%%%%

\section{The full Bayesian approach to LR}\label{expo}
The likelihood ratio assessment often involves some unknown nuisance parameters, denoted as $\theta$. In our case, it is the proportion of individuals in the relevant population with Y-STR profile corresponding to that of the matching trace, or the entire vector containing the frequencies of all the profiles.
The parameter of interest, $h$, is the unknown true hypothesis. 
Available data is made of evidence ($E$) directly related to the crime, and which helps in discriminate $h$, and additional background data ($B$) not directly related to the crime and only pertaining to the nuisance parameter $\theta$. This is partially different from the `background information' $I$ as defined in \citet{aitken:2004, taroni:2014}, but we can say that often background data can be thought of as part of the background information.

The difference between Bayesian and frequentist methods consists in how they treat the parameters $\theta$ and $h$. A Bayesian models the uncertainty about their value by random variables $\Theta$ and $H$, which are given prior distributions $p(\theta)$ and $p(h)$. Frequentists consider them as fixed (i.e., without distribution) unknown quantities. The reader is invited to notice the difference between $\theta$ and $h$: one is the parameter which we `test' through the likelihood ratio ($h$), the other ($\theta$) is a nuisance parameter involved in its calculation. 
%Bayesians model $h$ through the random variable $H$ with values  $\{h_p, h_d\}$. The prior given to $H$ is irrelevant for the LR assessment.
Some assumptions about the conditional independence probability for the model can be made, valid both for the frequentist and for the Bayesian approach. 
\begin{description}
\item{\textbf{a.}} The distribution of $B$ given $h$ and $\theta$, only depends on $\theta$.
\item{\textbf{b.}} $B$ is independent of $E$, given $\theta$ and $h$.
\end{description}
In our DNA example, condition \textbf{a} corresponds to ask that the sampling mechanism to obtain the database of reference is independent of which hypothesis is correct. This is true if, as it often happens, the database is collected before the crime.
Condition \textbf{b} holds if the suspect has been found on a ground of different evidence that has nothing to do with DNA.
In what follows we are going to use Bayesian network notation to specify the conditional independence relations of the proposed models. We expect the reader to be familiar with such a representation. 
\subsection{Bayesian point of view.}\label{bayes}

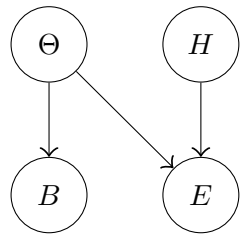
\begin{figure}[htbp]
\begin{center}
\begin{tikzpicture}
  \node[draw, circle, minimum size=1cm]                (a) at (0,0)  { $\Theta$ };
  \node [draw, circle, minimum size=1cm]                  (b) at (2,0)  { $H$ };
  \node [draw, circle, minimum size=1cm]                      (c) at (2,-2) { $E$};
  \node [draw, circle, minimum size=1cm]              (d) at (0,-2) { $B$};
   \draw[black, big arrow] (a) -- (c);
 \draw[black, big arrow] (b) -- (c);
 \draw[black, big arrow] (a) -- (d);
\end{tikzpicture}
\caption{Bayesian network representing the dependency relations between $E$ (evidence of the case) $B$ (background data in the form of a database) $\Theta$ (population parameter) and $H$ (hypotheses of interest).}\label{bnet_t1}

\end{center}
\end{figure}

Bayesians deal with the uncertainty over the parameters $\theta$ and $h$ by considering their values as realisations of, respectively, random variables $\Theta$ and $H$.  
A full Bayesian model is defined when the prior joint probability distribution for all the random variables of the model (here $E$, $B$, $H$ and $\Theta$) is given. 
This full Bayesian model can be thus represented by the Bayesian network of Figure~\ref{bnet_t1}, which is in turn equivalent to the following three conditions:
\begin{description}
\item{\textbf{Bayesian a.}} $B$ is conditionally independent of $H$ given $\Theta$.
\item{\textbf{Bayesian b.}} $B$ is conditionally independent of $E$ given $\Theta$ and $H$.
\item{\textbf{Bayesian c.}} $\Theta$ is unconditionally independent of $H$.
\end{description}

Notice that they are the Bayesian reformulation of conditions \textbf{a.} and \textbf{b.} mentioned above, with an additional condition (\textbf{Bayesian c.}) which corresponds to assuming that the Bayesian probability makes $\Theta$ and $H$ independent, and 
 is guaranteed for instance if prior beliefs on $\theta$ and on $h$ are assessed by people with different responsibilities and tasks: a judge for $h$ and a DNA expert (or a statistician) for $\theta$. However, notice that by definition the LR is independent of the prior belief over $h$.

The structure of the Bayesian network (or, equivalently, the three conditions above) allows to factorise the joint prior as $p(\theta, h, b, e) = p(\theta)p(h) p(b|\theta) p(e|\theta, h).$
The Bayesian probability $\Pr$ underlying the model is defined accordingly. As all Bayesian probabilities it is an expression of the subjective belief of the experts. This is achieved by choosing the prior distribution for $\theta$ and $h$ which reflects expert's beliefs. The distribution of all other variables given $\theta$ and $h$ is defined by the model, and need no subjective assessment.

The Bayesian likelihood ratio can be derived in the following way:
\begin{align*}
\LR=&\frac{\Pr(E=e, B=b|H=h_p)}{\Pr(E=e, B=e|H=h_d)}=\frac{\Pr(E=e|B=b, H=h_p)}{\Pr(E=e| B=b, H=h_d)}
= \frac{\int p(e | b, h_p, \theta)\, p(\theta | b, h_p) \text{d}\theta}{\int p(e | b, h_d, \theta)p(\theta | b, h_d) \text{d}\theta }
\\=&  \frac{\int \theta\, p(\theta| b)\, \text{d}\theta}{\int \theta^2\, p(\theta| b)\, \text{d}\theta}
= \frac{\mathbb{E}(\Theta|B=b)}{\mathbb{E}(\Theta^2|B=b)}.
\end{align*}
where some simplifications have been carried out because of conditions \textbf{a}, \textbf{b}, and \textbf{c}. Moreover, the first equality is due to the fact that it is implied by the network structure that $B$ is also unconditionally independent of $H$, and the second to last equality is due to the fact that, given condition \textbf{a} it holds that $p(\theta|b, h)=p(\theta|b), \,  \forall h$.

\subsection{Frequentist  point of view.}
As already mentioned, frequentists consider $h$ and $\theta$ as fixed quantities, whose unknown values correspond to, respectively, the true value of $\theta$ and the correct hypothesis.
The frequentist model can be thus seen as a special case of the Bayesian model described in Section~\ref{bayes}, where $\Theta$ and $H$ are given degenerate priors on $\theta$ and $h$, respectively. Alternatively, one can express the frequentist probability $\mathpzc{Pr}$ in terms of the Bayesian $\Pr$ in the following way: 
$\mathpzc{Pr}(\cdot):= \mathpzc{Pr}^{\theta}_h(\cdot)= \Pr(\cdot | H=h, \Theta=\theta).
$
If the Bayesian $\Pr$ was subjective, the frequentist $\mathpzc{Pr}$ is a measure which is universally defined by Nature.

Regarding $h$, according to prosecution its true value is $h_p$, while according to defence it is $h_d$. So one can think of two different frequentist probabilities: one for the prosecution ($\mathpzc{Pr}^{\theta}_{h_p}$) and one for the defence ($\mathpzc{Pr}^{\theta}_{h_d}$).

From a frequentist point of view, conditions \textbf{a} and \textbf{b} correspond to ask that:
\begin{description}
\item{\textbf{Frequentist a.}} $\mathpzc{Pr}^{\theta}_{h_p}(B=b)=\mathpzc{Pr}^{\theta}_{h_d}(B=b)$, for all $\theta$ and $b$.
\item{\textbf{Frequentist b.}} $\mathpzc{Pr}^{\theta}_{h}(E=e|B=b)=\mathpzc{Pr}^{\theta}_{h}(E=e)$, for all $\theta, h, e$, and $b$.
\end{description}
Obviously, \textbf{Bayesian c} becomes irrelevant in the frequentist framework.

The frequentist $\mathpzc{LR}$ can be derived as:
\begin{equation}\label{eq22x}
\mathpzc{LR}= \frac{\mathpzc{Pr}^{\theta}_{h_p}(E=e, B=b)}{\mathpzc{Pr}^{\theta}_{h_d}(E=e, B=b)}= \frac{\mathpzc{Pr}_{h_p}^\theta(E=e\mid B=b)\mathpzc{Pr}_{h_p}^\theta( B=b)}{\mathpzc{Pr}_{h_d}^\theta(E=e\mid B=b)\mathpzc{Pr}_{h_d}^\theta( B=b)}= \frac{\mathpzc{Pr}_{h_p}^\theta(E=e)}{\mathpzc{Pr}_{h_d}^\theta(E=e)}
\end{equation}
where the last equality is due to conditions \textbf{Frequentist a}, and \textbf{Frequentist b}.

\textcolor{black}{
Stated otherwise, frequentists look at a value for LR$|\theta$ (read ``LR given $\theta$''), where the value $\theta$ is fixed and has to be estimated through data.} 

Through observations, frequentists attempt to get close to the true $\mathpzc{LR}$ by choosing some estimator $\widehat{\mathpzc{LR}}$. One possibility is to 
estimate $\theta$ with a particular $\widehat{\theta}$. This leads to the so-called \emph{plug-in estimation} $\widehat{\mathpzc{LR}}=\mathpzc{LR}(\widehat{\theta})$ of the $\mathpzc{LR}$. However, that's not the only option \citep{cereda:2015b}.

By looking at \eqref{eq22x} the reader will realise that, if the frequentist approach is chosen, and under conditions \textbf{a} and \textbf{b}, one would get to the same result by evaluating only $E$ or both $E$ and $B$. This means that part of the information, namely $B$, is not useful to discriminate between the two hypotheses of interest (however, it usually plays an important role to obtain the estimate $\widehat{\theta}$ to be plugged into the $\mathpzc{LR}$). The same does not hold in the Bayesian context.

\subsection{The Bayesian plug-in LR and the proper Bayesian LR}

\begin{figure}[htbp]
\begin{center}
\begin{tikzpicture}
  \node[draw, circle, minimum size=1cm]                (a) at (0,0)  { $\Theta$ };
  \node [draw, circle, minimum size=1cm]                  (b) at (6,0)  { $H$ };
  \node [draw, circle, minimum size=1cm]                      (c) at (6,-2) { $E$};
  \node [draw, circle, minimum size=1cm]              (d) at (0,-2) { $B$};

  \draw[black, big arrow] (b) -- (c);
  \draw[black, big arrow] (a) -- (d);
\end{tikzpicture}
\caption{The two phase approach corresponding to Bayesian plug-in.}\label{bnet_t156}

\end{center}
\end{figure}
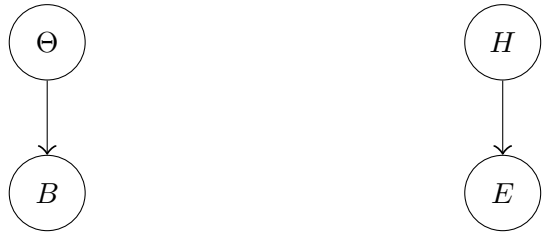

It is now time to discuss the fact that important forensic literature \citep[e.g.][]{evett:1998, balding:2005, lucy:2005} considers the likelihood ratio as `a measure of the probative value of the evidence regarding the two hypotheses' $h_p$ and $h_d$. According to this, it indicates the extent to which $E$ (and only $E$) is in favour of one hypothesis over the other. 
This is, in my opinion, the first important problem, since all data at disposal (namely $E$ and $B$) should be evaluated. Even though this is irrelevant in the frequentist framework (see \eqref{eq22x}), in the Bayesian framework for this definition to be appropriate we need to replace the probability $\Pr$ with the posterior probability $\Pr^*(\cdot)=\Pr(\cdot \mid B=b)$. Indeed, it holds that 
\begin{equation*} \label{LRw}
\text{LR}=\frac{\Pr(E=e, B=b \mid H=h_p)}{\Pr(E=e, B=b \mid H=h_d)}=\frac{\Pr(E=e \mid B=b, H=h_p)}{\Pr(E=e \mid B=b, H=h_d)}=\frac{\Pr^*(E=e \mid H=h_p)}{\Pr^*(E=e \mid H=h_d)}.
\end{equation*}
It is as if we have separated the evaluation process in two steps: first we observe $B=b$, and update the probability $\Pr(\cdot)$ to the posterior $\Pr^*(\cdot)=\Pr(\cdot \mid B=b)$, and then we define the likelihood ratio as the ratio of the probabilities ($\Pr^*$) of observing (only) the evidence $E$, under the two alternative hypotheses.\footnote{ Often, in literature \citep[][e.g.]{taroni:2014}, it is explicitly stated that $I$, the so-called background information, is omitted in the notation. We then agree with this choice provided that $B$ is part of $I$.}
This point is generally mistaken in literature and, as a consequence, the problem is split into two phases: first, a Bayesian estimate of $\theta$ using $B$, in the form of a posterior expectation is obtained, and then this estimate is plugged into a frequentist defined $\mathpzc{LR}$. 
It is as if, instead of using a combined model such as that in Figure \ref{bnet_t1}, they use two separate models as those in Figure~\ref{bnet_t156}: the left one is used to update the prior over the parameter.
The second one is used to derive the likelihood ratio (with $\theta$ considered as a fixed quantity).
In the end, they replace $\theta$ with the posterior expectation of $\Theta$ (now modelled through a random) given $B$.
This method will be referred to in the paper as the `Bayesian plug-in method',
since it is wrongly considered Bayesian, but it actually plugs in Bayes estimates into likelihood ratio defined in a frequentist way.

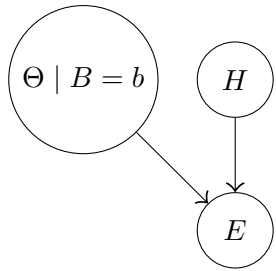
\begin{figure}[htbp]
\begin{center}
\begin{tikzpicture}
  \node[draw, circle, minimum size=1cm]                (a) at (0,0)  { $\Theta \mid B=b$ };
  \node [draw, circle, minimum size=1cm]                  (b) at (2,0)  { $H$ };
  \node [draw, circle, minimum size=1cm]                      (c) at (2,-2) { $E$};

  \draw[black, big arrow] (a) -- (c);
  
  \draw[black, big arrow] (b) -- (c);

\end{tikzpicture}
\caption{Updated Bayesian network after the observation of $B$.}\label{bnet_t16}

\end{center}
\end{figure}

\textcolor{black}{The correct Bayesian approach would be either to evaluate both $E$ and $B$ simultaneously, using the network of Figure~\ref{bnet_t1}, or in two steps: after the observation of $B$, we can update the model to the one represented in Figure~\ref{bnet_t16}, and use this for the evaluation of $E$.
}

\subsection{State of the art for DNA match evaluation}
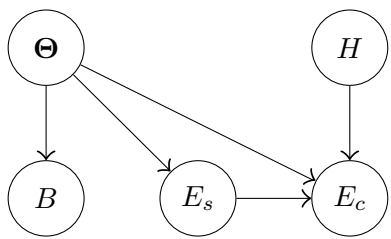
\begin{figure}[htbp]
\begin{center}
\begin{tikzpicture}
  \node[draw, circle, minimum size=1cm]                (a) at (0,0)  { $\boldsymbol{\Theta}$ };
  \node [draw, circle, minimum size=1cm]                  (b) at (4,0)  { $H$ };
  \node [draw, circle, minimum size=1cm]                      (c) at (2,-2) { $E_s$};
 \node [draw, circle, minimum size=1cm]                      (c2) at (4,-2) { $E_c$}; 
  \node [draw, circle, minimum size=1cm]              (d) at (0,-2) { $B$};
  \draw[big arrow] (a) -- (c);
    \draw[big arrow] (a) -- (c2);
  \draw[big arrow] (b) -- (c2);
    \draw[big arrow] (c) -- (c2);
  \draw[big arrow] (a) -- (d);
\end{tikzpicture}
\caption{Bayesian network for the DNA example.}\label{bnet_DNA}

\end{center}
\end{figure}

In case of a DNA match, we can use the Bayesian network of Figure~\ref{bnet_DNA}, which is equivalent to the network in Figure~\ref{bnet_t1} with the only difference that here node $E$ is split into two separated nodes, $E_s$ and $E_c$ representing the suspect's and the crime stain's profile, respectively.
We denote with $\boldsymbol{\theta}$ the unknown vector made of the population proportions of the different DNA profiles in the population, modelled through the random variable $\Theta$. Here we assume that we know the whole list of different DNA types present in the population of possible perpetrators, later we will consider the situation in which we don't. With $\Theta_{e_s}$ we will denote the population frequency of the suspect's (and crime stain's) profile.

 According to the prosecution, the suspect left the stain. This implies that $\Pr(E_c=e_s|\boldsymbol{\Theta}=\boldsymbol{\theta}, E_s= e_s,H= h_p)=1$, under the assumption that each true match is correctly reported. According to the defence, another person from the population left the stain, hence the probability of it being exactly of type $e_c$ is equal to the population proportion of that profile: $\Pr(E_c=e_s|\boldsymbol{\Theta}=\boldsymbol{\theta}, E_s= e_s,H= h_d)=\theta_{e_s}$.
Moreover, it holds that $p(b|e_s, \boldsymbol{\theta})\Pr(e_s|\boldsymbol{\theta})p(\boldsymbol{\theta})$ is proportional to $p(\boldsymbol{\theta}| e_s, b)$. The correct Bayesian procedure would lead to:
\begin{align*}
\LR=&\frac{\Pr(E=e,B=b|H=h_p)}{\Pr(E=e,B=b|H=h_d)}
=\frac{\int\Pr(E_c=e_s|H=h_p,\boldsymbol{\Theta}=\boldsymbol{\theta}, E_s=e_s)p(e_s|\boldsymbol{\theta}) p(b|\boldsymbol{\theta}, e_s)p(\boldsymbol{\theta})d\boldsymbol{\theta}}{\int\Pr(E_c=e_s|H=h_d,\boldsymbol{\Theta}=\boldsymbol{\theta}, E_s=e_s)p(e_s|\boldsymbol{\theta}) p(b|\boldsymbol{\theta}, e_s)p(\boldsymbol{\theta})d\boldsymbol{\theta}}\\
=&\frac{\int p(\boldsymbol{\theta}|e_s, b)d\theta}{\int \theta_{e_s} p(\boldsymbol{\theta}|e_s, b)d\theta}=\frac{1}{\mathbb{E}(\Theta_{e_s}|E_s=e_s,B= b)}.
 \end{align*}

On the other hand, the common approach taken by forensic literature would be to propose the following derivation for the likelihood ratio
\begin{align*}
\LR=&\frac{\Pr(E=e|B=b, H=h_p)}{\Pr(E=e|B=b, H=h_d)}=\frac{\Pr(E_c=e_s|B=b, E_s=e_s,H=h_p)\cancel{p(e_s|b,h_p)}}{\Pr(E_c=e_s|B=b, E_s=e_s,H=h_d)\cancel{p(e_s|b, h_d)}}\\
=&\frac{1}{\Pr(E_c=e_s|H=h_d)}=\frac{1}{\theta_{e_s}}.
 \end{align*}
 Then, $\theta_{e_s}$ is replaced with $\widehat{\theta_{e_s}}=\mathbb{E}(\theta_{e_s}|B=b)$.
 
 Let us focus on the second to last equality. By looking at the Bayesian network we could already realise that actually $E_c$ is independent only if $\theta$ is given. Thus, it is not allowed to simplified $E_s$ in the conditioning unless we are considering frequentist probabilities ($\mathpzc{Pr}$).
 Last equality is also incorrect. Indeed, it holds that $\Pr(E_c=e_s\mid B=b, E_s=e_s, h_d)= \int \Pr(E_c=e_s \mid \boldsymbol{\Theta}=\boldsymbol{\theta}, E_s=e_s, h_d)p(\boldsymbol{\theta}\mid b, e_s, h_d)\text{d}\boldsymbol{\theta}$.
 
 It is true that, in the end, computationally, the difference amounts on using $\mathbb{E}(\theta_{e_s}|B=b, E_s=e_s )$ instead of $\mathbb{E}(\theta_{e_s}|B=b)$ (i.e., the well-known problem of whether to add or not the suspect to the database before the posterior) and thus the plug-in can be seen as an approximation of the full Bayesian approach. However, it is an hybrid solution, thus conceptually ill defined.

This hybrid approach is often considered Bayesian since the lack of knowledge about $\theta$ is dealt with using Bayesian posterior mean $\widehat{\theta_{e_s}}=\mathbf{E}(\Theta_{e_s}|B)$ as a point estimate of $\theta_{e_s}$ \citep{weir:1996, curran:2005, taroni:2010, sjerps:2015}. This is why we will refer to this way of proceeding as the \emph{Bayesian plug-in method}. 
 As pointed out in \citet{weir:1996}, ``either the mean or the mode of the posterior distribution can serve as an estimate but each is merely a summary of the whole distribution". 
Not only this method is hybrid and inconsistent, but it suffers from several weaknesses. For instance, one would obtain different $\widehat{\mathpzc{LR}}$s depending on whether one wants to estimate $\theta_{e_s}$, $1/\theta_{e_s}$ or $\log_{10}(1/\theta_{e_s})$: this arbitrariness is in some way entailed in the idea of `estimating' the likelihood ratio. 
Moreover, as stated in \citet{taroni:2015}, the likelihood ratio (meaning the Bayesian one) should be calculated, rather than estimated. Including $B$ as part of the data to evaluate, and applying simple Bayesian theory, we can calculate the Bayesian LR, without any estimation
needed.
Notice that already \citet{foreman:1997} and \citet{brummer:2014} proposed a differentiation between the `plug-in estimates' and the `full Bayesian analysis'.
%
%\textcolor{red}{SPOSTARE QUI NON SERVONO A NULLA, accennare al problema nella intro ma non qui A conventional choice for the prior distribution of the haplotype frequency $\theta$ is the beta distribution. Sometimes, the whole set ($p_1,..., p_k$) is chosen as $\theta$. In that case the Dirichlet distribution is proposed as a prior for allelic frequencies from multiallelic loci \citep{good:1965, lange:1995, weir:1996, taroni:2010}.
%Both this choices suffers from a problem: the models and the priors are defined after seeing the data, and this violates Bayesian principles. }
%\textcolor{red}{For instance, in the Beta- binomial model, the random variable $\Theta$ models the frequency of the DNA type of the crime scene. Thus its definition and its prior is chosen after the data are provided. In the Dirichlet-multinomial model parameter $\Theta$ is a vector containing the different DNA profiles in the population. Typically, which types and how many of them is unknown. If they are chosen according to the types in $B$ then the same mistake is done. A solution may be that or randomize these two quantities, as proposed in Section~\ref{diri}
%} .

\section{A useful Lemma}\label{lem}

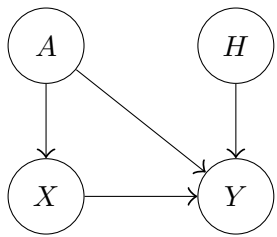
\begin{figure}[htbp]
  \begin{center}
  \begin{tikzpicture}
\node[draw, circle, minimum size=1cm]                (t) at (-1,0)  { $A$};
\node [draw, circle, minimum size=1cm]              (h) at (1.5,0) {$H$};
 \node [draw, circle, minimum size=1cm]              (d) at (-1,-2) { $X$};
\node [draw, circle, minimum size=1cm]              (dr) at (1.5,-2) { $Y$};
  \draw[black, big arrow]  (t) -- (d);
  \draw[black, big arrow] (h) -- (dr);     
  \draw[black, big arrow]  (t) -- (dr);
   \draw[black, big arrow] (d) -- (dr);
\end{tikzpicture}

\end{center}
     \caption{Conditional dependencies of the random variables of Lemma\ref{lemma1}}
\label{figure1gggg}
     \end{figure}

Lemma~\ref{lemma1} is a result regarding four general random variables $A$, $X$, $Y$, $H$ whose conditional dependencies are represented by the Bayesian network of Figure~\ref{figure1gggg}. This is important due to the possibility of applying it to a very common forensic situation: the prosecution and the defence disagree on the the distribution of part of data ($Y$) but agree on the distribution of the other part ($X$). The distribution of $X$ and $Y$ depends on some parameter(s) modeled by $A$.

   \begin{lemma}\label{lemma1}

 Given four random variables $A$, $H$, $X$ and $Y$, whose conditional dependencies are represented by the Bayesian network of Figure~\ref{figure1gggg}, the likelihood function for $h$, given $X=x$ and $Y=y$ satisfies 
$$\mathrm{lik}(h\mid x, y) \propto \mathbb{E}(p(y \mid x, A, h) \mid X = x).$$

 \end{lemma}

 A proof of this lemma can be found in \citet{cereda:2015b}. We will see an application of it in Sections~\ref{beta} and \ref{diri}.

\section{Bayesian LR calculation, based on beta-binomial model}\label{beta}
In the binomial model, the database of size $N$ is regarded as the result of a sequence of $N$ Bernoulli trials with parameter $\theta$, where \emph{success} corresponds to the observation of the same type of that observed at the crime scene, and failure to the observation of any other type. Let's denote by $b$ the number of successes among these $N$ experiments. 
When data is treated as a binomial outcome, the most conventional choice of the prior for the parameter $\theta$ (probability of success) is the beta distribution, due to the famous conjugacy property. 
In forensic and medical statistics literature there are many examples for the use of this distribution for a genetic (autosomal) frequency \citep{weir:1996,gunel:1995, roewer:2000, brenner:2010,  buckleton:2011, biedermann:2008b, biedermann:2013b}. 
$$\Theta \sim \text{Beta}(\alpha, \beta).$$
The observation of the suspect's profile $E_s$ corresponds to another Bernoulli trial, a successful one in the case of interest (the suspect matches the crime stain type). 
The information provided by the database and the suspect's type can be reduced to the count of profiles of this type in this sample of size $N+1$ (database and suspect) from the population of interest. 
%In fact, our sample sequence can be regarded as `exchangeable', meaning that the probability of observing $b+1$ successes is the same regardless of the order in which these successes appear. In other words, $b+1$ and $N$ carry all information that can be obtained regarding $\theta$ from the sample \citep{good:1965}. Because of exchangeability, the binomial distribution provides a sensible model, when the data arises from a sequence of draws from a large population.  
$$B \mid \Theta=\theta \sim \text{Bin}(N, \theta)$$
$$B, E_s\mid \Theta=\theta \sim \text{Bin}(N+1, \theta)$$
Notice that according to the defence $E_c$ can be seen another Bernoulli experiment of the same kind. On the other hand, according to the prosecution it is equal to 1 with probability one. 
Stated otherwise, $$E_c \mid E_s=1, H=h\sim \begin{cases} \delta_1 & \text{if }H=h_p\\
\theta & \text{if } H=h_d \end{cases}.$$

The Bayesian network of Figure~\ref{bnet_DNA} can be used for this model. Hence, we can apply the Lemma~\ref{lemma1} using $X=(B, E_s)$ (part of data whose distribution is agreed on by defence and distribution) and $Y=E_c$ (part of data whose distribution is disagreed on by defence and distribution). 
The LR can thus be developed in the following way:\begin{align}\label{Lrbeta}
\text{LR}=&\frac{\mathbb{E}(\Pr(E_c=1| E_s=1, H=h_p, \Theta) \mid E_s=1, B=b) }{\mathbb{E}(\Pr(E_c=1|E_s=1, H=h_d, \Theta) \mid E_s=1, B=b )}
=\frac{1}{\mathbb{E}(\Theta \mid E_s=1, B=b )}=\frac{\alpha + \beta + N+1}{\alpha+b +1}.
\end{align}
The last equality is due to the fact that, using the well known beta binomial conjugacy property, it holds that
$$\Theta \mid B=b, E_s=1 \sim \text{Beta}(\alpha +b+1,\beta + N - b).$$ 
The LR as in \eqref{Lrbeta}, also proposed in \citet{dawid:1996} and \citet{taroni:2015}, can be compared to the one obtained with the `standard' Bayesian plug-in estimate \citep{weir:1996, taroni:2010}:
\begin{equation*}
\widehat{\LR}= \frac{\alpha+\beta+N}{ \alpha+b}.
\end{equation*}
It is easy to see that the Bayesian plug-in $\widehat{\LR}$ is a non conservative estimate of LR, in a way that is unfavourable to the defence. Indeed, $\LR< \widehat{\LR} \Leftrightarrow \beta+N>b$, which is always true, since $b\leq N$ and $\beta>0$.
\textcolor{black}{Notice that there is an alternative derivation for \eqref{Lrbeta}. It can be obtained in a two step evaluation: first, the observation of the database $B$ and of the suspect haplotype $E_s$ updates the probability $\Pr$, then the updated probability is used to calculate the LR for the observation of another identical haplotype (the one found at the crime scene).}
\textcolor{black}{\begin{description}
\item[First step] The probability $\Pr$ is updated to $\Pr^{**}(\cdot)=\Pr(\cdot|E_s=1, B=b)$ after the database and the haplotype of the suspect are observed. In practice, the prior distribution Beta($\alpha, \beta$) on $\theta$ is updated to the posterior Beta($\alpha+1, \beta+N-1$). 
\item[Second step] The new probability $\Pr^{**}$ is used to calculate the likelihood ratio for the observation of the haplotype from the crime scene:  $$\LR=\frac{\Pr^{**}(E_c=1|H=h_p)}{\Pr^{**}(E_c=1|H=h_d)}= \frac{1}{\mathbb{E^{**}}(\Theta)}=\frac{1}{\frac{\alpha+b+1}{\alpha+\beta+1+N}}.$$
\end{description}}

\begin{figure}
\centering
\begin{minipage}{.33\textwidth}
  \centering
  \includegraphics[width=.99\linewidth]{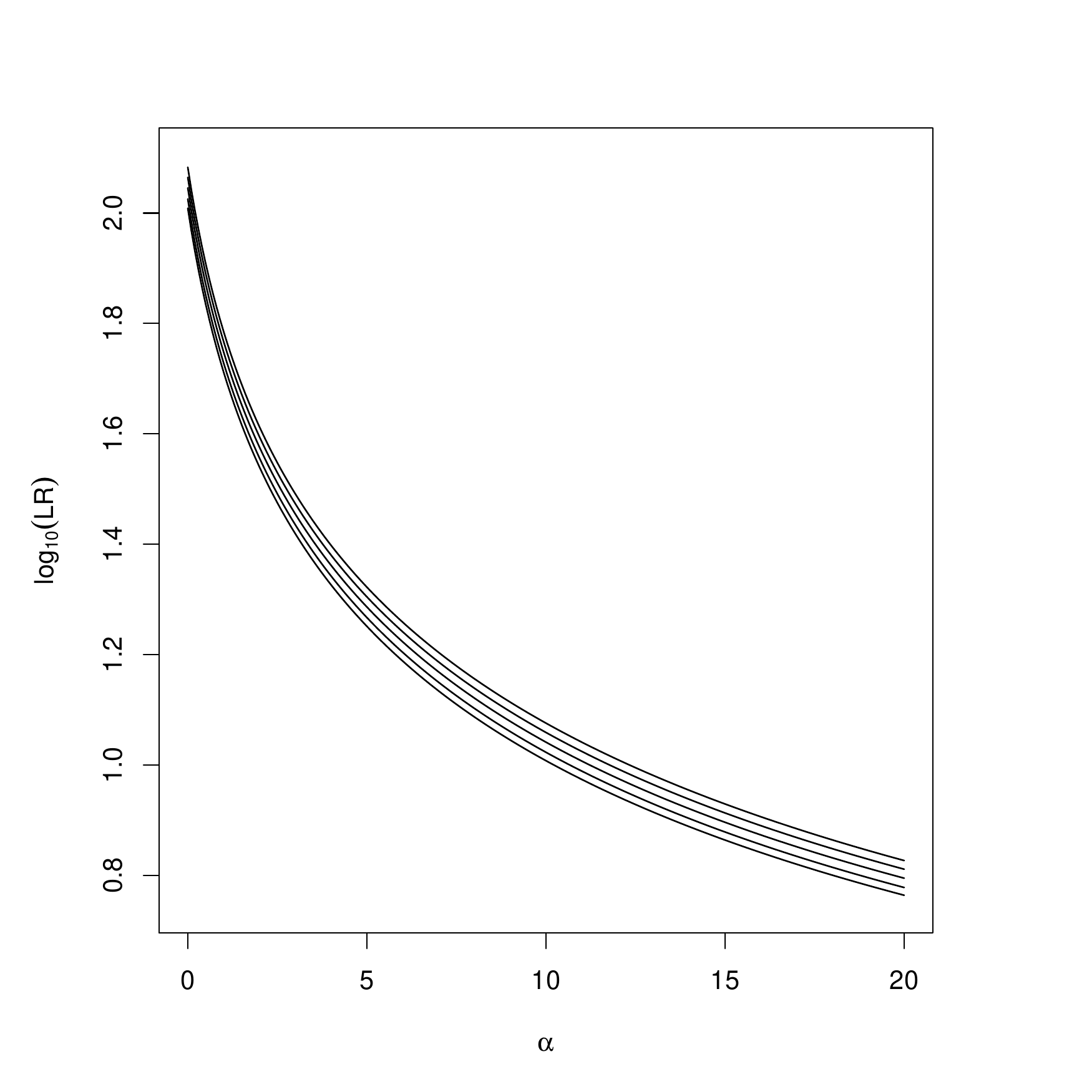}
  \captionof{subfigure}{$\log_{10}\LR$}  \label{fig:test1}
\end{minipage}%
\begin{minipage}{.33\textwidth}
  \centering
  \includegraphics[width=.99\linewidth]{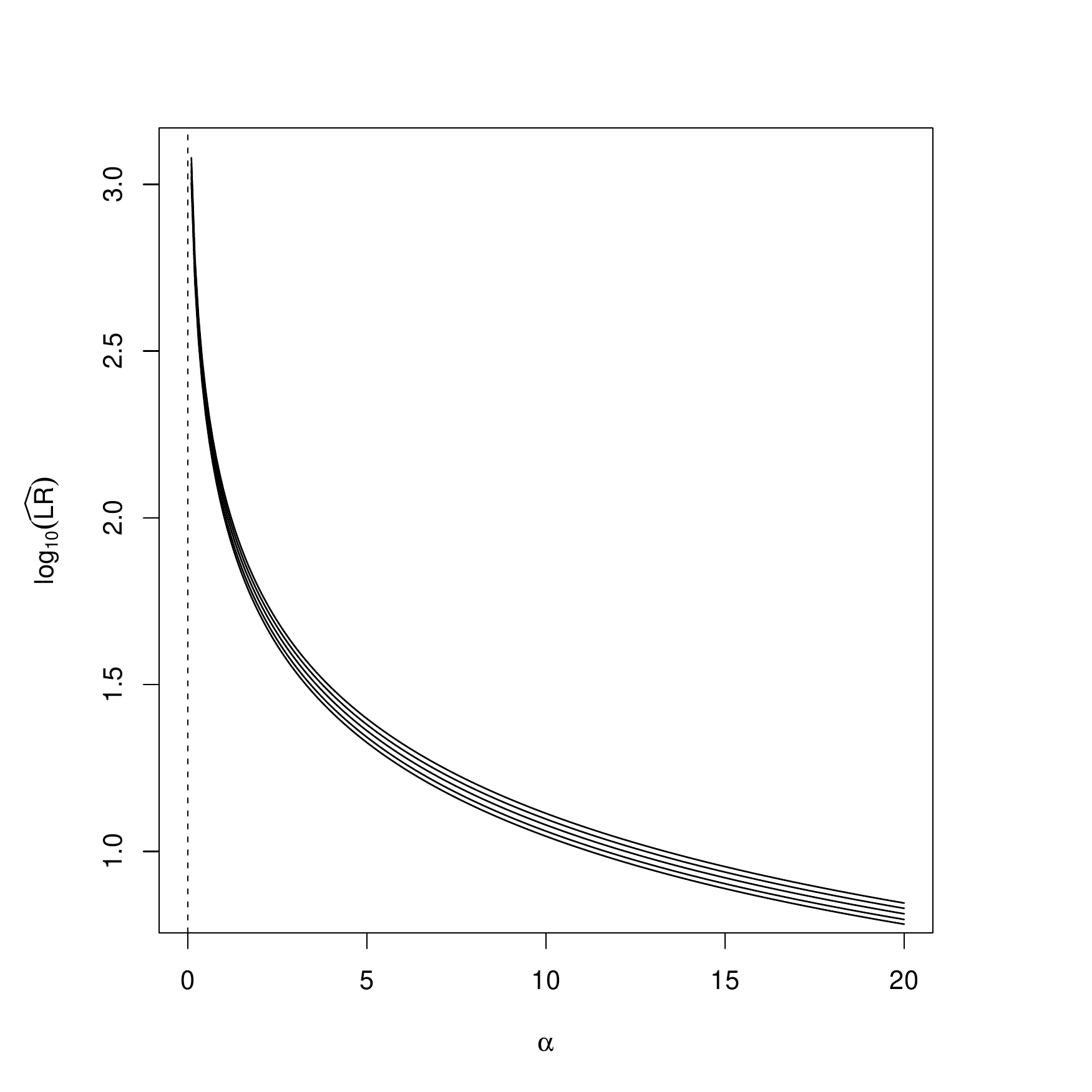}
 \captionof{subfigure}{$\log_{10}\widehat{\LR}$}  \label{fig:test2}
\end{minipage}
\begin{minipage}{.33\textwidth}
  \centering
  \includegraphics[width=.99\linewidth]{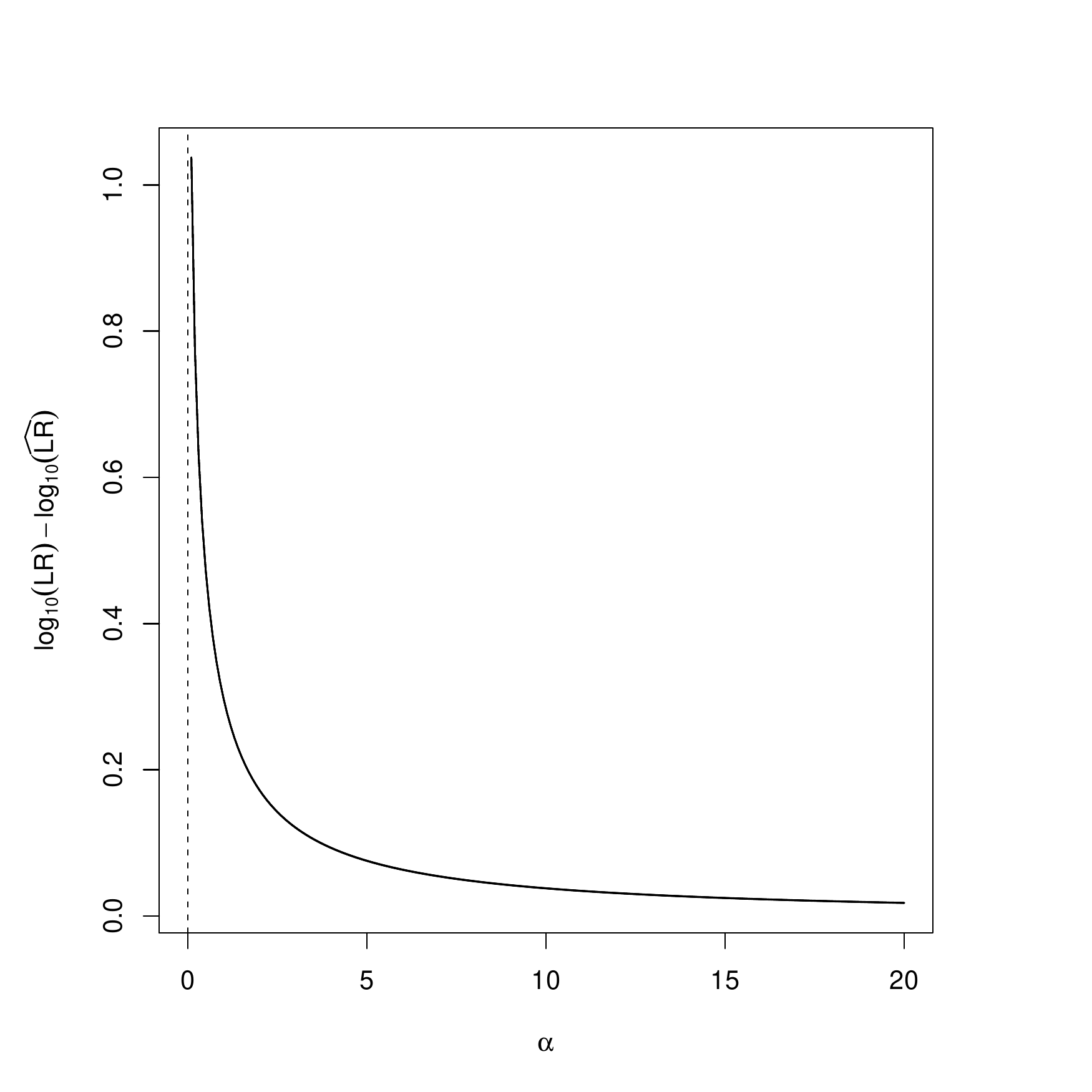}
 \captionof{subfigure}{$\log_{10}\widehat{\LR} - \log_{10}\LR$ }  \label{fig:test3}
\end{minipage}
\captionof{figure}{Sensitivity analysis for the three quantities $\log_{10}\LR$ (a),  $\log_{10}\widehat{\LR}$ (b),  $\log_{10}\widehat{\LR} - \log_{10}\LR$ (c), in the beta binomial model, when $\alpha \in (0, 20]$ (x axis), and for $\beta\in\{1,5,10,15,20\}$ (corresponding to the different lines, where highest line corresponds to highest $\beta$).}   \label{fig1200}
\end{figure}

\paragraph{Sensitivity analysis.}
The sensitivity of the quantities $\log_{10}\LR$, $\log_{10}\widehat{\LR}$, and of the difference between them, to the hyperparameters $\alpha$ and $\beta$ of the beta prior is shown in Figure~\ref{fig1200}, for the rare type case (i.e., $b$ = 0), and with $N=100$. In particular, the figure shows the variation of $\log_{10}\LR$ (a), of the plug-in estimate $\widehat{\log_{10}\LR}=\log_{10}\widehat{\LR}$ (b), and of the difference $\log_{10}\widehat{\LR}- \log_{10}\LR$ (c), when different values of $\alpha$ (x axis) and $\beta$ (only five values corresponding to the different lines) are chosen in the interval $(0, 20]$.

%
% \begin{figure}[htbp]
% 
%\medskip
%\hspace{0.65\baselineskip}\hfil
%\makebox[0.25\textwidth]{(a) $\log_{10}\LR$}\hfil
%\makebox[0.25\textwidth]{(b) $\log_{10}\widehat{\LR}$}\hfil
%\makebox[0.25\textwidth]{(c)  $\log_{10}\widehat{\LR}-\log_{10}\LR$}
%
%
%\settoheight{\tempdim}{\includegraphics[width=0.9\textwidth]{Figure_1.pdf}}%
%
%\includegraphics[trim=0cm 0cm  0cm 0cm, clip=true,width=0.33\textwidth]{Figure_1b.pdf}\hfil
%\includegraphics[trim=0cm 0cm 0cm 0cm, clip=true,width=0.33\textwidth]{Figure_2b.pdf}\hfil
%\includegraphics[trim=0cm 0cm 0cm 0cm, clip=true,width=0.33\textwidth]{Figure_3b.pdf}
%
% 
%
%\caption{Contour plots representing the values of $\log_{10}\LR$ (a),  $\log_{10}\widehat{\LR}$ (b),  $\log_{10}\widehat{\LR} - \log_{10}\LR$ (c) for the beta binomial model when $\alpha,\beta \in (0.01, 2)$, $b=0$, and $N=100$. }\label{fig1200}
%\end{figure}

Observing Figure~\ref{fig1200} (or analysing \eqref{Lrbeta}), it can be seen that the three quantities of interest hardly depend on $\beta$, while they decrease as $\alpha$ increases. 
In particular, when $\alpha$ decreases to 0, $\log_{10}\LR$ behaves as $\log_{10}(1+\beta+N)$, while $\log_{10}\widehat{\LR}$ increases to $+\infty$. Another way to see this is that, for fixed $\beta$, as $\alpha$ increases, the prior distribution on $\theta$ resembles more and more to the degenerate distribution localised on the value $\theta=1$ (notice this is inappropriate for the rare type match case). 
This means that the haplotype whose population proportion is modelled through the random variable $\Theta$ (i.e., the haplotype of the crime stain and of the suspect) has probability one to be observed, which leads to $\widehat{\LR}=1$ (hence, $\log_{10}\LR=0$). 
On the other hand, if $\alpha$ decreases to zero, the prior distribution over $\theta$ tends to resemble to the degenerate distribution localised on the value $\theta=0$. This leads to  $\widehat{\LR}=1/0=+ \infty$. On the whole, the plug-in estimate $\widehat{\LR}$ is less stable than $\LR$, as can be seen comparing Figures~\ref{fig1200} (a) and (b), in the sense that is more sensitive to changes in $\alpha$ (especially for small values).
The difference, represented in (c) has, for fixed $\beta$, a vertical asymptote when $\alpha\rightarrow 0$, increasing as fast as $\log_{10}1/\alpha$. On the other hand it decreases to $0$ with an horizontal asymptote when $\alpha\rightarrow \infty$.  
 From Figure~\ref{fig1200} (c) it can be observed that the difference is important only for small values of $\alpha$. Otherwise the two methods would lead essentially to the same conclusions, so that the plug-in can be seen as a good approximation of the proper Bayesian procedure.
\section{Bayesian LR calculation, based on Dirichlet-multinomial model}\label{diri}
\begin{figure}[htbp]
\begin{center}
\begin{tikzpicture}
  \node[draw, circle, minimum size=1cm]                (a) at (-2,0)  {$ \boldsymbol{\Theta}$};
  \node [draw, circle, minimum size=1cm]                  (b) at (2,0)  { $H $};
  \node [draw, circle, minimum size=1cm]                      (c) at (2,-2) {$ E_c$};
  \node [draw, circle, minimum size=1cm]                      (c2) at (0,-2) {$ E_s$};
  \node [draw, circle, minimum size=1cm]              (d) at (-2,-2) { $B$};
   \node[draw, circle, minimum size=1cm]                (e) at (-2,2)  { \textbf{Type}};
    \node[draw, circle, minimum size=1cm]                (f) at (-2,4)  { $K$};
      \draw[black, big arrow] (a) -- (c2);
  \draw[black, big arrow]  (a) -- (c);
     \draw[black, big arrow]  (c2) -- (c);

   \draw[black, big arrow] (b) -- (c);
  \draw[black, big arrow] (a) -- (d);
  \draw[black, big arrow] (f) -- (e);
       \draw[black, big arrow] (e) -- (a);
\end{tikzpicture}
\caption{Bayesian network for Dirichlet-multinomial model, when $k$ is randomized. }\label{bnet_ch}
\end{center}
\end{figure}
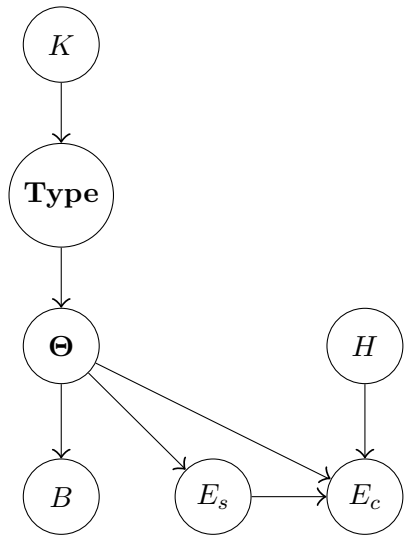

When database is treated as a multinomial sample of size $N$ from a population with $k$ different haplotypes, the conventional choice of the prior for the vector  $\boldsymbol{\theta}$ containing the population frequencies of all the different haplotypes in Nature is the Dirichlet distribution. 
 Literature provides many examples of the use of this method for the frequencies of autosomal markers \citep[][e.g.]{curran:2002,balding:1995b, lange:1995, weir:1996,buckleton:2005b, taroni:2010}. However, these methods don't consider the uncertainty about the number $k$ of possible types in the population, and this can be a problem especially since we want to apply it to Y-STR haplotypes, for which the database often does not offer a good coverage. If, in addition, we are using the model for the rare type match case, then we have to find a solution.
 The problem of estimating $k$ is a very challenging one. It has been addressed both with frequentist methods \citep[][e.g.]{ chao:1992, haas:1998} and with Bayesian methods \citep[][e.g.]{hill:1968, lewins:1984, barger:2010}.  
We propose the derivation of a full Bayesian LR which uses priors over the number $k$ of different types in the population. The model is represented by the Bayesian network of Figure~\ref{bnet_ch}. The bottom part (from node $\boldsymbol{\Theta}$ down) has a well-known structure (see Figure~\ref{bnet_DNA}), while the upper part needs further explanation.

Assume that there may be at most $m$ theoretically possible profiles alphabetically\footnote{Remember each profile is a list of numbers.} ordered in a vector, called $\mathbf{s}$. For instance, $m=20^{10}$ (10 loci, with 20 possible alleles each). 
Only $k$ of them are actually present in Nature (or more specifically in the population of interest), but $k$ is not known and also which of the $m$ are those $k$ is not known. 

We will denote as  $K$ the random variable which represents how many of the $m$ potentially possible haplotypes are actually present in the population of interest. The prior distribution for $k$ is denoted generically as $p(k)$. The random vector $\mathbf{Type}$, of length $k$, contains the ordered positions, in vector $\mathbf{s}$, of the $k$ haplotypes of the population of interest. A particular configuration of $\mathbf{Type}$ is denoted as $\mathbf{t}=(i_1, ..., i_k)$, where $i_1<...<i_k$. $\mathbf{t}$ is chosen uniformly at random from the possible $\binom{m}{k}$ configurations.
The random vector $\boldsymbol{\Theta}$ contains the population proportions of all the haplotypes, both those whose position is contained in $\mathbf{Type}$, and those that are not (corresponding to zero entries). A particular configuration of $\boldsymbol{\Theta}$ is denoted  $\mathbf{\theta}=(\theta_1, ..., \theta_{m}$), many entries of which are zero. We assume that the positive entries, i.e., $(\theta_i \mid i\in \mathbf{t})$, are drawn from a $k$ dimensional Dirichlet distribution with all $k$ hyper parameters $\alpha$ equal to $1$. 
%Different choices could have been made for the hyper parameters of the Dirichlet, but this is out of the scope of this section, which wants to show the full Bayesian derivation for the LR and compare it to the plug-in version.
Now, as usual, $H$ represents the hypotheses of interest, and can take the value $h \in \{h_p, h_d\}$, according to the prosecution or the defence, respectively. 
%The prior distribution for $H$ is again irrelevant for the LR assessment. 
$E_s$ and $E_c$ contain the index $e_s$ and $e_c$ of the haplotypes of the suspect and of the crime scene, respectively. In the situation of interest $e_c=e_s$.
Lastly, random vector $\mathbf{B}$ represents the database, seen as a multinomial sample from the population with parameters $N$ and $\boldsymbol{\theta}$. 
A particular configuration of $\mathbf{B}$ is denoted $\mathbf{b}= (b_1, ..., b_{m})$ representing the absolute frequency in the database of each of the $m$ haplotypes. It contains $k_{obs}<k$ positive values, and many zeros.
By applying Lemma~\ref{lemma1} to this situation we have that 
$$\LR=\frac{\mathbb{E}(\Pr(E_c=e_s\mid E_s=e_s, \mathbf{B}=\mathbf{b}, \boldsymbol{\Theta}, H=h_p)\mid E_s=e_s, \mathbf{B}=\mathbf{b})}{\mathbb{E}(\Pr(E_c=e_s\mid E_s=e_s, \mathbf{B}=\mathbf{b}, \boldsymbol{\Theta}, H=h_d)\mid E_s=e_s, \mathbf{B}=\mathbf{b})}=\frac{1}{\mathbb{E}(\Theta_{e_s}\mid E_s=e_s, \mathbf{B}=\mathbf{b})}.$$

It can be proven that for $\alpha=1$ this leads to

\begin{equation}\label{123r}\LR=\frac{1}{2} \frac{{\displaystyle \sum_{k=k_{obs}+1}^m \binom{k}{k_{obs}+1}\frac{\Gamma(k)}{\Gamma(k+N+1)}p(k)}}{{\displaystyle\sum_{k=k_{obs}+1}^m \binom{k}{k_{obs}+1} \frac{\Gamma(k)}{\Gamma(k+N+2)}p(k)}}
\end{equation}

Notice that the likelihood ratio depends on the data only through $k_{obs}$. This is due to the choice of the symmetric Dirichlet prior, and of the uniform prior for $\mathbf{Type}$. In particular, this tells us that data can be reduced by sufficiency to $k_{obs}$.
The likelihood ratio obtained through a classical plug-in Bayesian estimation is:
\begin{equation}\label{blabla}
\widehat{\LR}=\frac{\bar{k}\alpha+N}{\alpha+b_{e_s}}=\bar{k}+N .
\end{equation}
where the number of haplotypes is a fixed value $\bar{k}$, to be chosen (or estimated) in advance.
In order to compare the two values \eqref{blabla} and \eqref{123r} we need to choose a value for $\bar{k}$. A reasonable choice can be $\bar{k} = \mathbf{E}(K)$.
Among the possible priors one can choose for $K$, we decided to test the Poisson distribution (see Section~\ref{prior_poisson}) and the negative binomial distribution (see Section~\ref{prior_negbin}).

\subsection{Poisson prior}\label{prior_poisson}

In this section a Poisson distribution with parameter $\lambda$, truncated so as to have support only on $\{1,2,...,m\}$, is chosen as prior distribution for $K$.
$$p(k):= p(k;\lambda)\propto 
\begin{cases}
\frac{e^{-\lambda}\lambda^{k}}{k! }&  \textrm{if}\ k \in \{ 1,..., m\}\\
0&  \text{elsewhere}
\end{cases}$$
where $\lambda>0$. If $\lambda$ and $m$ are large enough, the normalising constant can be omitted and we have the standard Poisson distribution:

$$p(k;\lambda)=\frac{e^{-\lambda}\lambda^{k}}{k! }, \quad \forall k \in \mathbb{N}$$

The LR in~\eqref{123r} becomes

\begin{equation*}
\LR= \frac{1}{2} \frac{\sum_{k=k_{obs}+1}^{m} \frac{\lambda^k}{k-k_{obs}-1!}\frac{\Gamma(k)}{\Gamma(N+k+1)}}{\sum_{k=k_{obs}+1}^{m} \frac{\lambda^k}{k-k_{obs}-1!}\frac{\Gamma(k )}{\Gamma(N+k+2)}}
\end{equation*}

It is then of interest to analyse the quantities $\log_{10}\LR$, $\log_{10}\widehat{\LR}$, and the difference $\log_{10}\widehat{\LR} - \log_{10}\LR$ between them and to carry on a sensitivity analysis to see how these quantities vary when parameter $\lambda$ changes.

\paragraph{Sensitivity analysis}\label{a1}

\begin{figure}[htbp]
\centering

 \begin{minipage}{.45\textwidth}
  \centering
  \includegraphics[width=.9\linewidth]{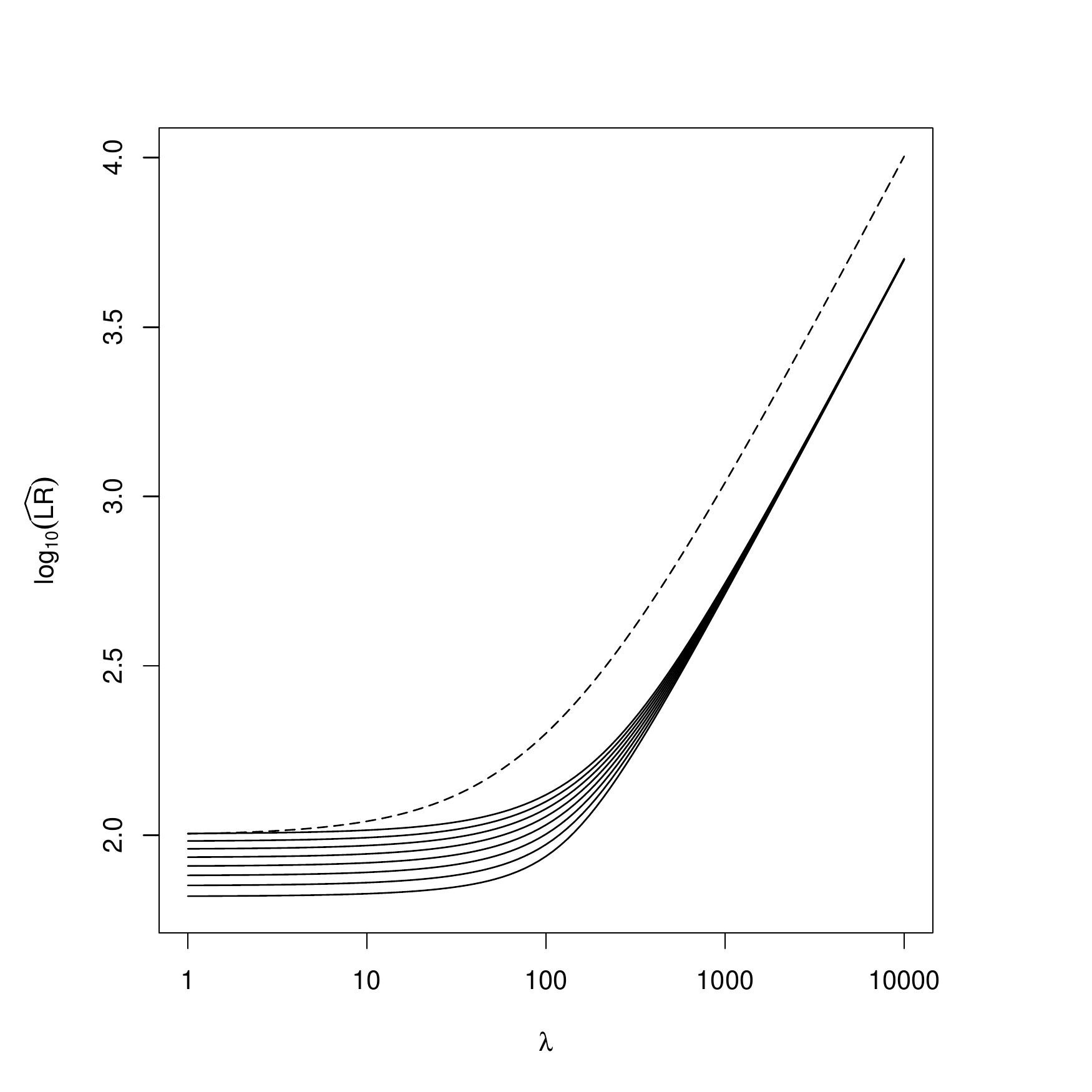}
 \captionof{subfigure}{$\log_{10}\widehat{\LR}$}  \label{fig66test2}
\end{minipage}
\begin{minipage}{.45\textwidth}
  \centering
  \includegraphics[width=.9\linewidth]{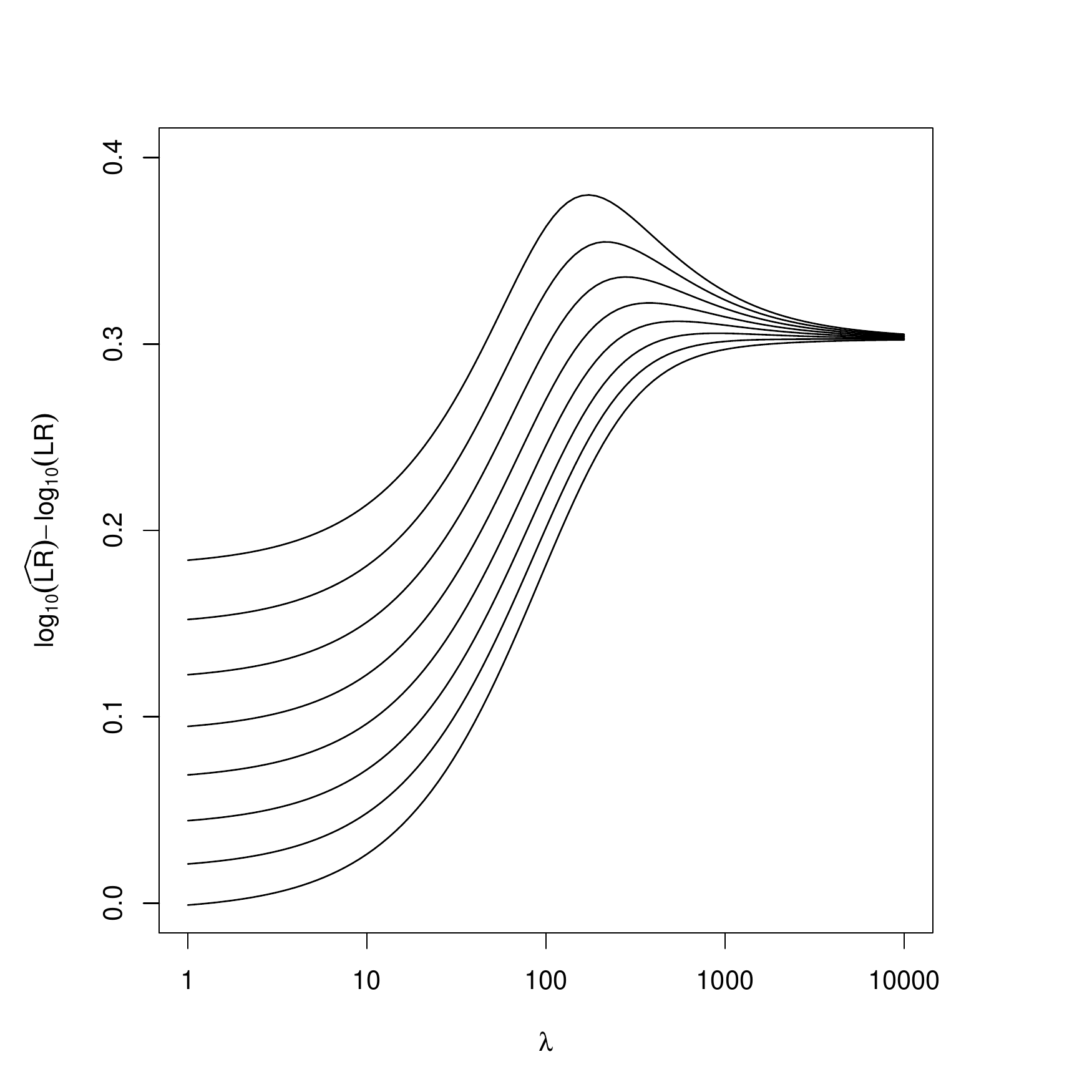}
 \captionof{subfigure}{$\log_{10}\widehat{\LR} - \log_{10}\LR$ }  \label{fig66test3}
\end{minipage}
\captionof{figure}{Poisson prior, for the Dirichlet model. Sensitivity analysis of $\log_{10}\LR$ (a, black lines), $\log_{10}\widehat{\LR}$ (a, dashed line), and of the difference $\log_{10}\widehat{\LR}-\log_{10}\LR$ (b), to different values of $\lambda \in [1, 10\,000]$ (x axis), and $k_{obs}\in\{30, 40, 50, 60, 70, 80, 90, 100\}$ (represented by the different lines where highest line corresponds to highest $\beta$).}   \label{figgtag2}
\end{figure}

In the rare type match problem (i.e., $b_{e_s}=0$), when a Poisson($\lambda$) prior is chosen for the dimension $K$ of the Dirichlet distribution (with all parameters $\alpha$ equal to 1), the sensitivity of the three quantities $\log_{10}\LR$, $\log_{10}\widehat{\LR}$, and of their difference, to $\lambda$ and $k_{obs}$ is shown in Figure~\ref{figgtag2}, when $N=100$. 

%\begin{figure}[htbp]
%\medskip
%\hspace{0.45\baselineskip}\hfil
%\makebox[0.3\textwidth]{(a) $\log_{10}\LR$}\hfil
%\makebox[0.33\textwidth]{(b) $\log_{10}\widehat{\LR}$}\hfil
%\makebox[0.33\textwidth]{(c)  $\log_{10}\widehat{\LR}-\log_{10}\LR$}
%
%
%\settoheight{\tempdim}{\includegraphics[width=0.4\textwidth]{Figure_1.pdf}}%
%
%\includegraphics[width=0.3\textwidth]{Figure_4b.pdf}\hfil
%\includegraphics[width=0.3\textwidth]{Figure_5b.pdf}
%\includegraphics[width=0.3\textwidth]{Figure_6b.pdf}
%
% 
%
%\caption{Plots of $\log_{10}\LR$ (a), $\log_{10}\widehat{\LR}$ (b) , and of the difference $\log_{10}\widehat{\LR}-\log_{10}\LR$ (c), for different values of $k_{obs}$ and $\lambda$, when $\alpha=1$, $b_{i_s}=0$, $N=100$, and a Poisson($\lambda$) prior is chosen for $K$.}\label{figgtag2}
%\end{figure}

%
%\begin{figure}[htbp]
%\begin{center}
%\includegraphics[width=0.4\textwidth]{Figure_10.pdf}
%
%\caption{ Behaviour of $\log_{10} \LR$, for $\lambda \in [10, 1000]$, when $b_{e_s}=0$, $N=100$, $k_{obs} = 70$, $\alpha = 1$, and a Poisson($\lambda$) prior is chosen for $K$.}\label{dds1}
%\end{center}
%\end{figure}
%

In particular, it can be inferred that the LR depends little on $k_{obs}$ and a lot on $\lambda$. 
When $\lambda$ is big (which is typically true $\lambda$ being the expected value of the number of different Y-STR haplotypes in a population) the LR depends almost only on $\lambda$. In particular, LR increases linearly with $\lambda$, since $\LR \sim  \lambda/2$.
This can be explained by replacing the Poisson prior on $k$, by the degenerate distribution localised on (the integer part of) $\lambda$: $f_K(k)=f(k;\lambda) =\mathds{1}_{\{\lambda\}}(k)$, for $\lambda \in\{1,2,....\}$. This approximation is sensible for large values of $\lambda$ in virtue of the law of large numbers (the Poisson($\lambda$) being the sum of $\lambda$ Poisson(1)). 
In this case \eqref{123r} becomes
\begin{equation*}
\LR= \frac{1+N+\lambda }{2} \sim \frac{ \lambda}{2}, \ \text{for} \ \lambda\rightarrow +\infty, \text{and} \  N \ \text{fix}.
\end{equation*}

 The plug-in estimates of $\log_{10}\widehat{\LR}$ (as defined in \eqref{blabla} and with the choice of $\bar{k}=\lambda$) is the dashed line shown in Figure~\ref{figgtag2} (a). The difference between the `true' value $\log_{10}\LR$, and the estimated one $\log_{10}\widehat{\LR}$  is shown in Figures~\ref{figgtag2} (b). In particular, one can see that, for big $\lambda$, it decreases when $\lambda$ increases and depends a little on $k_{obs}$, while for small values of $\lambda$ it has the opposite behaviour, and depends more strongly on $k_{obs}$. Note that, again, the plug-in method overestimates the LR by up to almost half of an order of magnitude. 
%\textcolor{blue}{The dependencies of the results on the hyper-parameters of the prior constitute, as for the beta binomial model, a bad characteristic of this model. }

%
%
\subsection{Negative binomial prior}\label{prior_negbin}

A different choice is that of using as prior for $k$ the negative binomial distribution \citep{hill:1968, hill:1979, lewins:1984} as prior distribution for $K$. For our model a negative binomial distribution truncated so as to have support $\{1, ... , m \}$ is more appropriate. It is defined as:
$$\Pr(K=k|r,q)\propto
\begin{cases}
\binom{k+r-1}{k}(1-q)^{k}q^{r}&  \textrm{if}\ k \in \{1, ..., m\}\\
0&   \text{elsewhere }
\end{cases}$$
where $r>0$ and $q\in(0,1)$. However, if $\mathbb{E}(K)$ is large, but small compared to $m$, the normalise factor is almost 1 and the standard negative binomial distribution can be used as prior distribution over $K$:
$$\Pr(K=k|r,q)= 
\binom{k+r-1}{k}(1-q)^{k}q^{r}, \quad \forall k \in \mathbb{N}.
$$

Using this prior, the likelihood ratio in \eqref{123r} becomes: 
\begin{equation} \label{a12a}
\LR=\frac{1}{2} \frac{\sum_{k=k_{obs}+1}^{m} (1-q)^k \frac{\Gamma(k )}{\Gamma(k  + N+1)}\frac{\Gamma(k+r)}{\Gamma(k-k_{obs})}}{\sum_{k=k_{obs}+1}^{m} (1-q)^k \frac{\Gamma(k )}{\Gamma(k  + N+2)}\frac{\Gamma(k+r)}{\Gamma(k-k_{obs})}}.
\end{equation}

In the following, a series of properties of the (zero truncated) negative binomial distribution will be listed, which will help to understand why this choice is more appropriate than the choice of the Poisson distribution as a prior for $K$. We will denote as NB($r,q$) a random variable distributed according to a negative binomial with parameters $r$ and $q$, and P($\lambda$) a random variable distributed according to a Poisson distribution with parameter $\lambda$.
\begin{enumerate}
\item The mean and variance of NB($r,q$) are, respectively, $\mathbf{E}$(NB($r,q$))$= (1-q)r/q$ and $ \mathbf{Var}$(NB($r,q$))$= (1-q)r/q^2$. This represents an advantage over the use of a Poisson distribution where these two quantities can't be tuned independently one another, since $\mathbf{E}(\text{P}(\lambda))=\mathbf{Var}(\text{P}(\lambda))=\lambda $. Thus, the use of a negative binomial prior guarantees more flexibility.
\item  The negative binomial NB($r,q$) is a Gamma mixture of Poisson.
\item For fixed $\lambda$=$\mathbf{E}(\text{NB}(r,q))$, when $r$ increases, the negative binomial $\text{NB}(r,q)$ tends in distribution to $\text{P}(\lambda)$. This means that the negative binomial distribution can be seen as an extension of the Poisson distribution.
\end{enumerate}
The same properties apply to the [$0,m$]-truncated case, both for the Negative Binomial, and for the Poisson, if $m$ is big enough \textcolor{red}{and the probability of 0 is small.}

\subsection{Sensitivity analysis}
A classical approach to sensitivity analysis for the negative binomial prior would be to analyse the sensitivity of $\log_{10}\LR$ to changes of $r$ and $q$, and $k_{obs}$, the three parameters appearing in \eqref{a12a}. However, we decided to use as parameters $r$, $k_{obs}$ (the number of different haplotypes observed in the database) and $\lambda$, the mean value of the negative binomial prior. In this way it is easier to see how the results depend on the average number of haplotypes in Nature, and that for big $r$ we fall back in the Poisson case, as explained in \textcolor{black}{property 3}.
Figure~\ref{figggag3} represents the sensitivity analysis for $\log_{10}\LR$, $\log_{10}\widehat{\LR}$ and the difference $\log_{10}\widehat{\LR}-\log_{10}\LR$ in the rare type case ($b_{i_s}=0$) for $\alpha=1$, $N=100$.

It can be inferred from this analysis that when $r$ increases the values depend more and more on $\lambda$ and less and less on $k_{obs}$. 

According to the second column of Figure~\ref{figggag3}, one can see that for $r \geq 100$ the plug-in estimate always exceeds $\log_{10}\LR$. Anyway, the difference is only significant if $r$ is small, in particular for high values of $\lambda$. 
\begin{table}[htbp]
  \centering
\renewcommand{\arraystretch}{0.008}
  \begin{tabular}{>{\centering\arraybackslash}m{0.59in} >{\centering\arraybackslash}m{2.5in} >{\centering\arraybackslash}m{2.5in}}
    				& 	(a) 				
				& 	(b) \\
    	$r=1$ 		&      \includegraphics[width=0.35\textwidth]{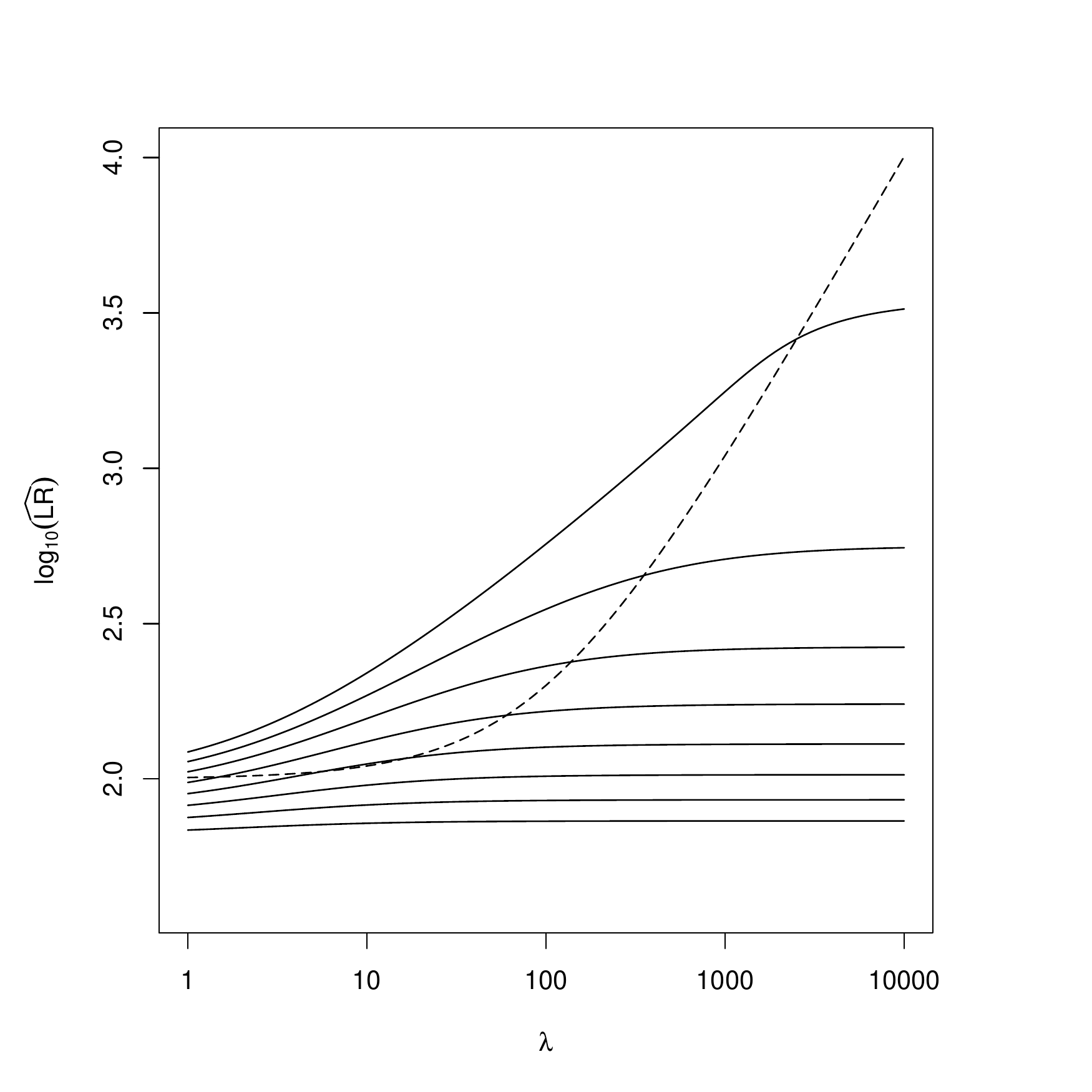}
      				& 	\includegraphics[width=0.35\textwidth]{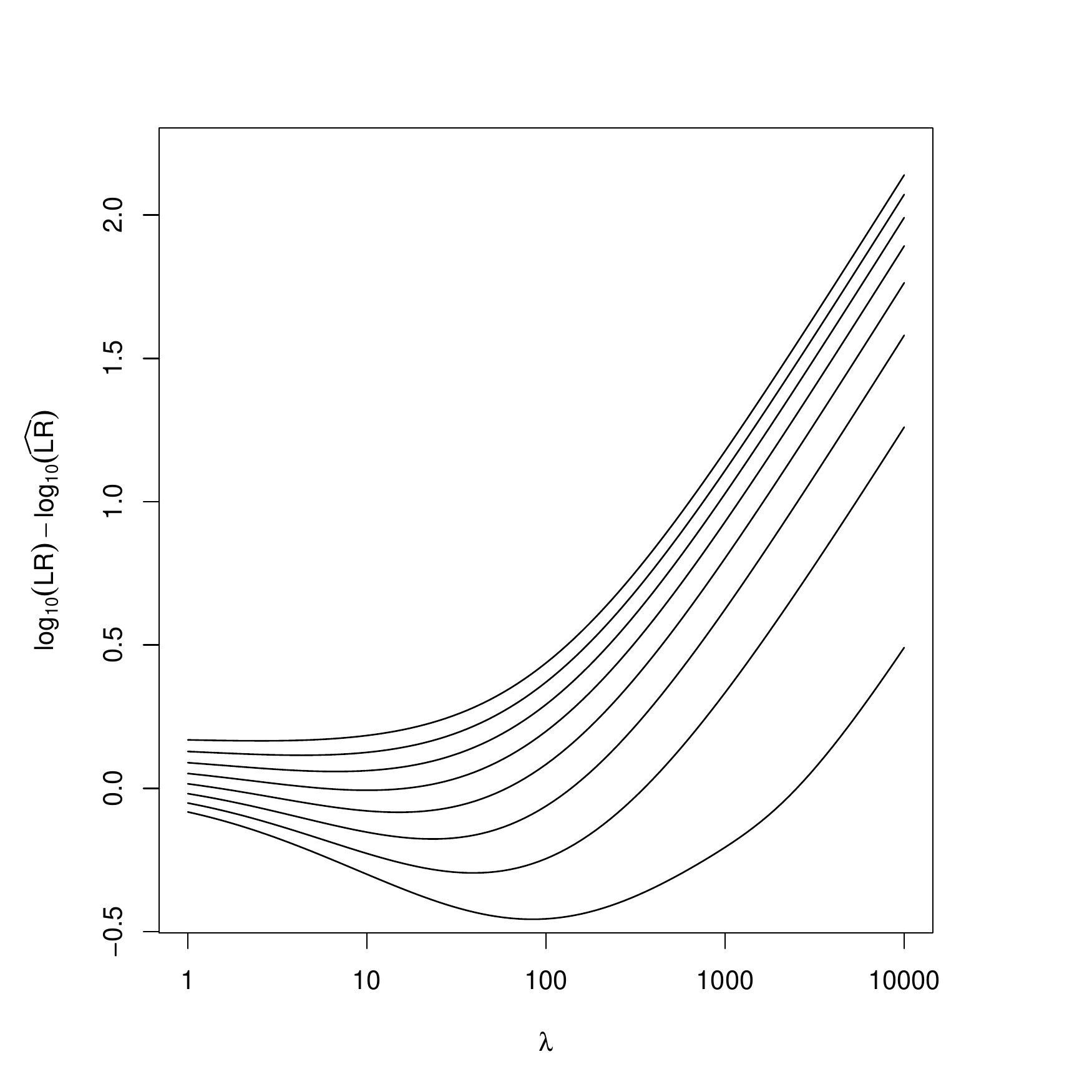}
\\ \hline
 	$r=10$ 		&      \includegraphics[width=0.35\textwidth]{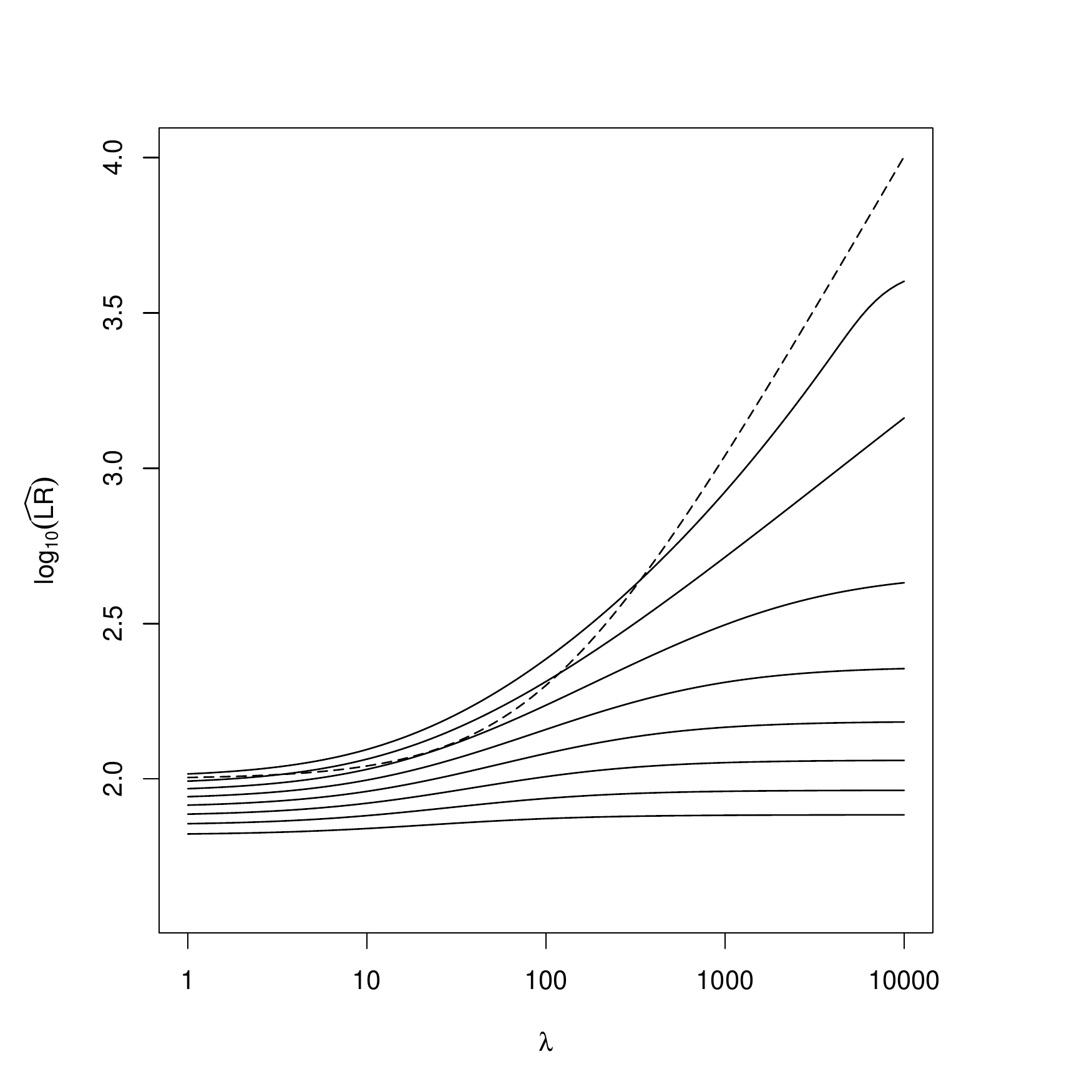}
        				& 	\includegraphics[width=0.35\textwidth]{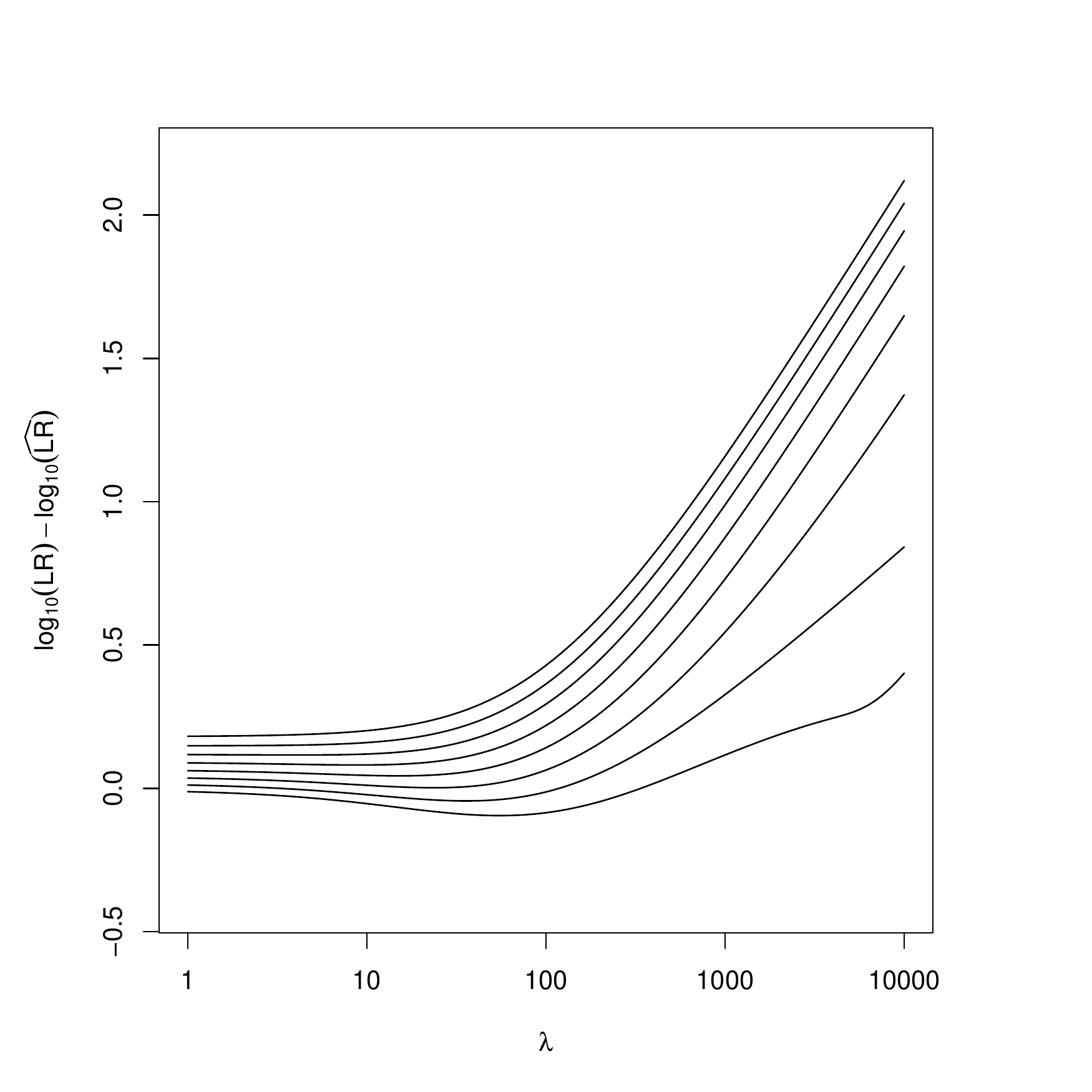}
\\ \hline
  	$r=100$ 		&	\includegraphics[width=0.35\textwidth]{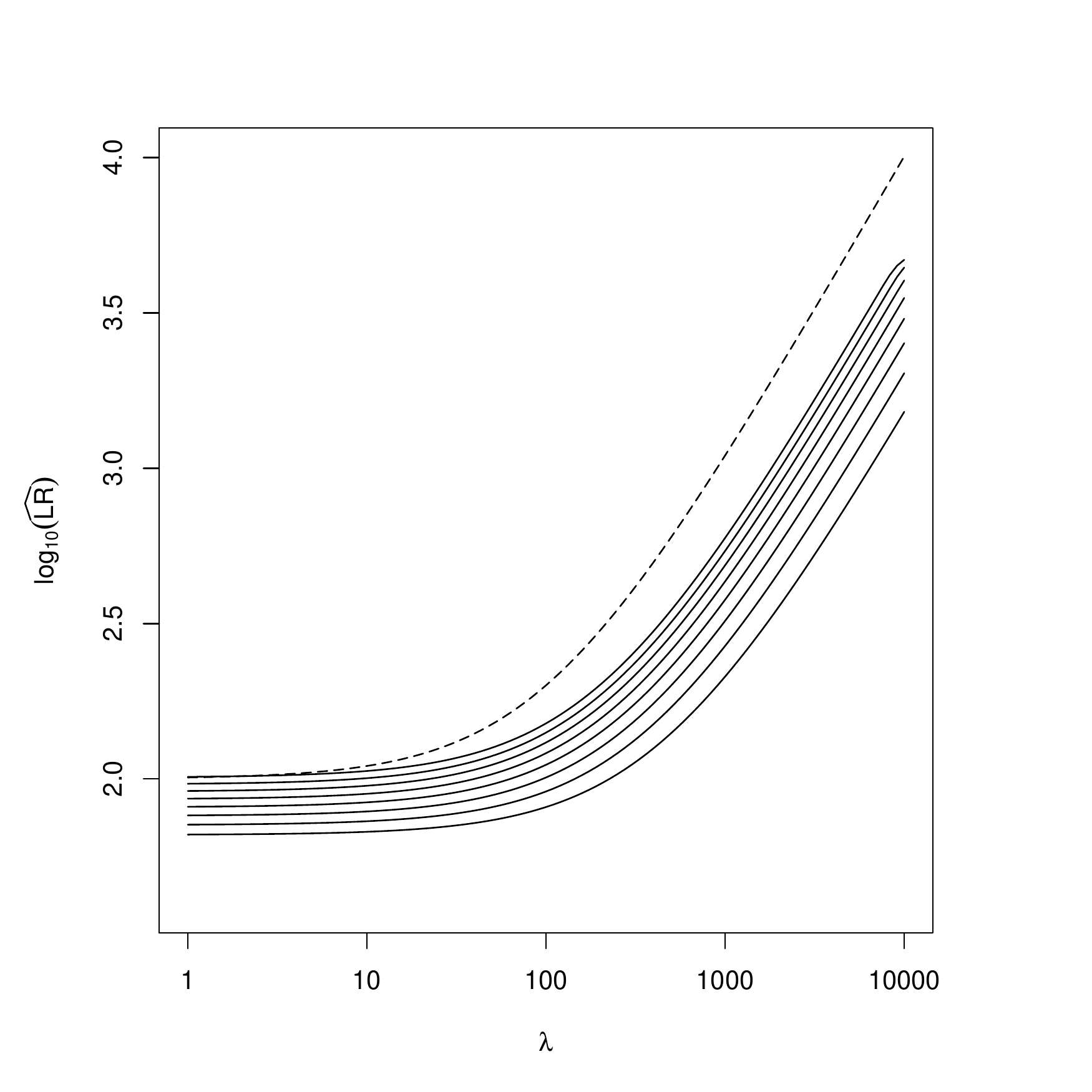}
      				&	\includegraphics[width=0.35\textwidth]{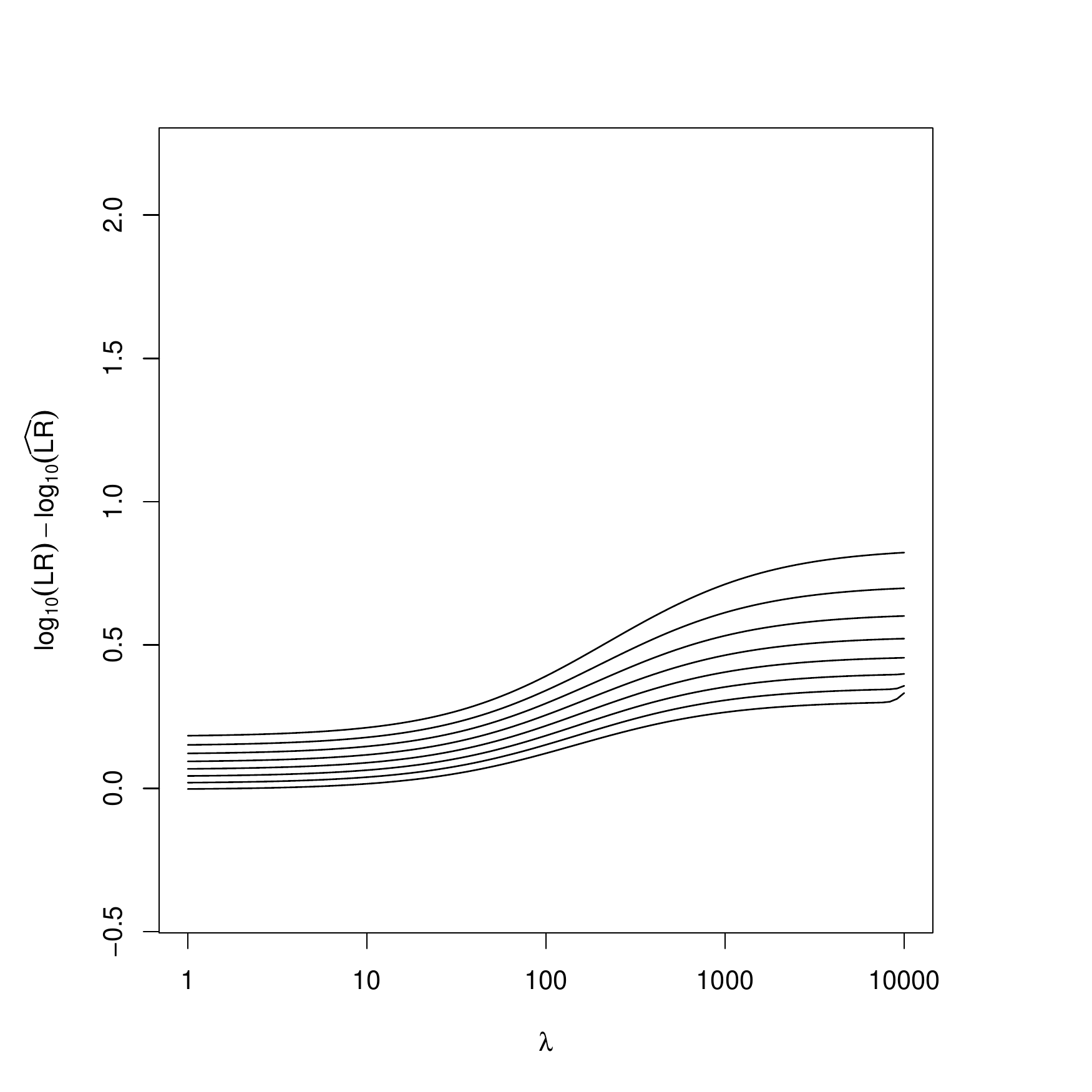}
 \\ \hline
 	$r=1000$ 		&	\includegraphics[width=0.35\textwidth]{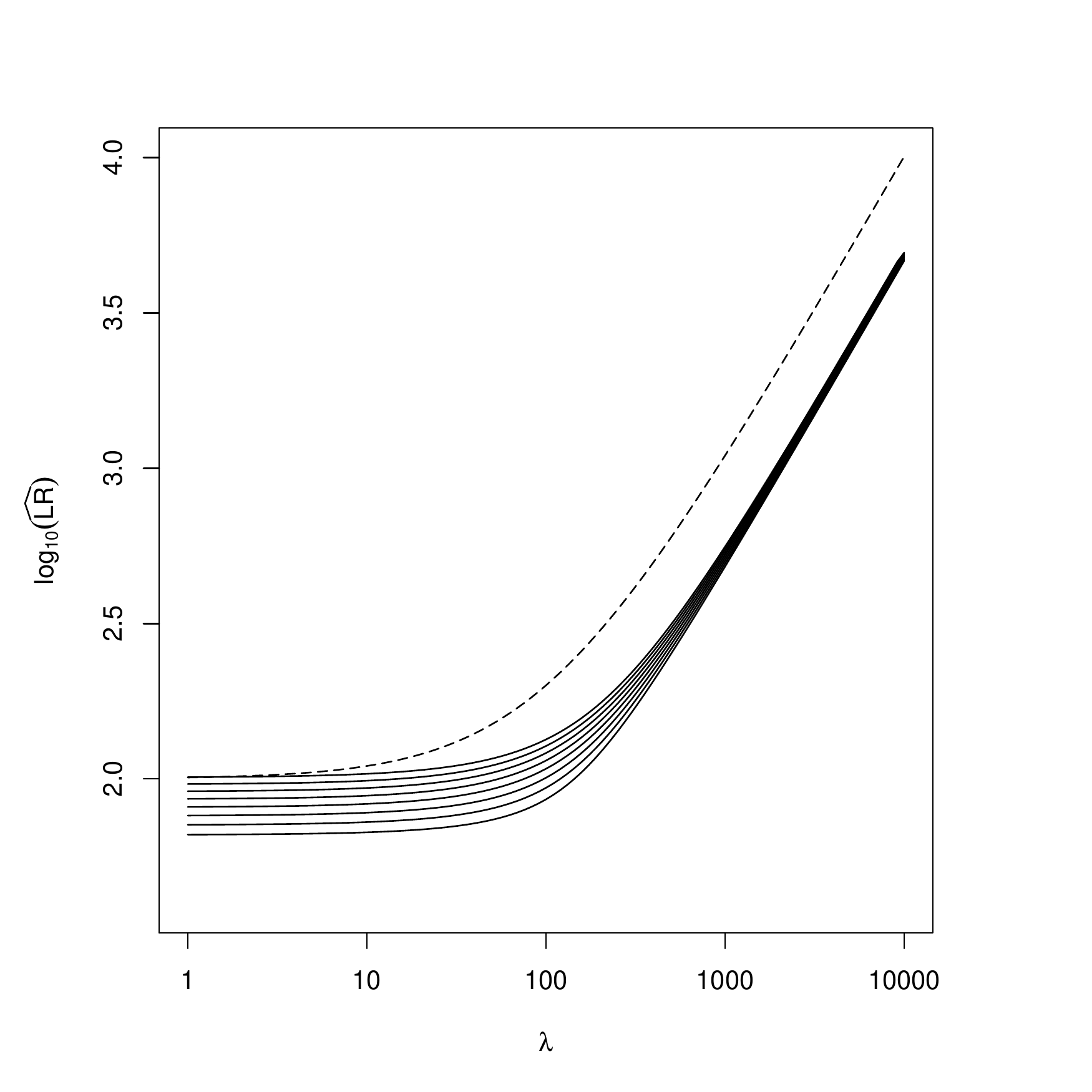}
     				& 	\includegraphics[width=0.35\textwidth]{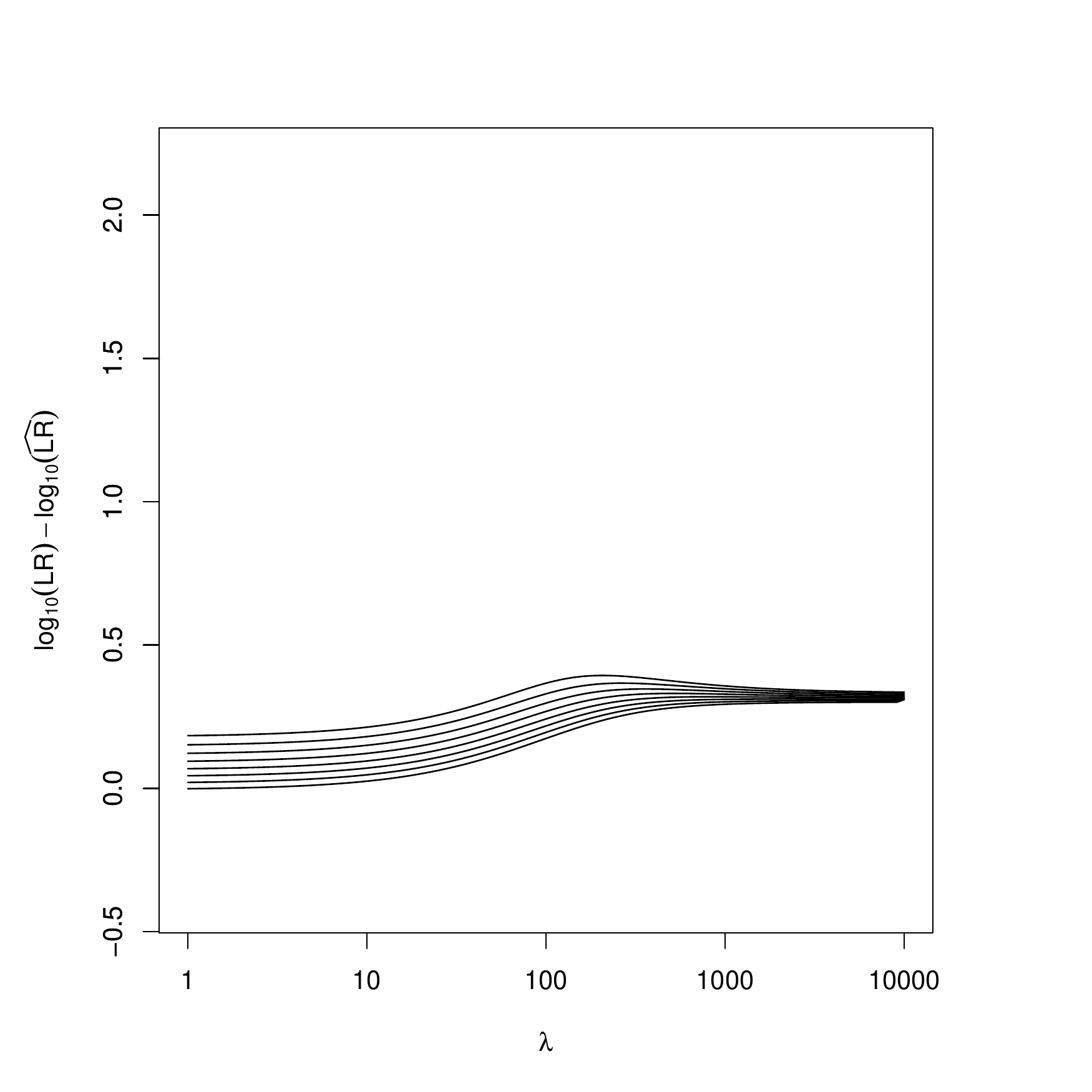}
%\\ 
%    $r=10000$ & \includegraphics[trim=0cm 0cm 0.5cm 1.5cm, clip=true,width=0.3\textwidth]{Figure_19b.pdf}
% 
%    &\includegraphics[trim=0cm 0cm 0.5cm 1.5cm, clip=true,width=0.3\textwidth]{Figure_20b.pdf}
%
  \end{tabular}
  \caption{Sensitivity analysis for the three quantities $\log_{10}\LR$ (first column, black lines), $\log_{10}\widehat{\LR}$ (first column, dashed line) and the difference $\log_{10}\widehat{\LR}-\log_{10}\LR$ (second column) to different values of $\lambda=\mathbf{E}(K)$ (x axis) and of $k_{obs}\in \{30, 40, 50, 60, 70 , 80, 90, 100\}$ (represented by the different lines, where highest line corresponds to the highest $k_{obs}$).}\label{figggag3}
\end{table}
\subsection{Remarks about conventional priors}
As mentioned above, the beta distribution and the Dirichlet distribution are the conventional choices for the prior in case of binomial or multinomial model, respectively. As stated in \citet{curran:2002}, this `remains the accepted standard in some laboratories''  because of the ``appeal of simplicity and ease of implementation''.  Although we agree that this may have been a very sensible reason some decades ago, nowadays,  with the computational skill provided by our computers, there are no more excuses to limit ourselves to these convenient priors. 
Indeed, a prior should reflect the expert beliefs rather than standards of computational ease. 

For the beta prior, the dependencies of the LR results on the value of the hyper parameter $\alpha$ stresses once more the need of a different choice. Moreover, the model is profile-specific, meaning that the beta priors is supposed to model the frequency of the profile observed at the crime scene. The model is thus to be defined after the data have been observed, and this seems to contradict common Bayesian principles.

For the Dirichlet prior, we have a similar issue. The dimension $k$ of this prior should correspond to the number of different DNA types in the population. This problem, which for autosomal markers could be easily overcome, is more important when Y-STR haplotypes are considered, the state space being huge, and the database hardly representative. If we choose as $k$ the number of different types observed in the database, then we are in trouble every time a new haploytpe is observed, as for the situation of interest for this paper. By treating $k$ as a Bayesian would do for an unknown quantity, we expected the likelihood ratio to depend a lot on the mean value of the prior chosen for $k$. 
The Dirichlet method with all parameters $\alpha=1$, and with a prior over $K$, turned out to depend only on the number of observed haplotypes in the database (and not on their frequencies). This is actually unattractive for Y-STR data, and is due to the symmetry. The data does not overrule the prior which makes \emph{all} the positive $p_i$ the same in size, and it is also the reason why the likelihood ratios obtained using the two methods (beta-binomial, and Dirichlet-multinomial) do not differ too much. Notice that for this prior we only focused on the case in which all the parameters $\alpha$ are equal to 1.
More could have been done, for instance explore the sensitivity of the likelihood ratio to changes in the $\alpha$ \citep{triggs:2006}, or use hierarchical model \citep{chen:2008}. However, we preferred to investigate other types of prior \citep{cereda:2015c} which we believe are more appropriate for Y-STR haplotypes frequencies.

The two methods of Section~\ref{beta} and Section~\ref{diri} differ in the choice of information retained 
from the database. The Beta method only retains as information the frequency of the observed haplotype. A lot of information regarding other haplotypes is discarded, such as how many have been observed, and their frequencies. Let us point out that if there will ever be guidelines on how to choose the hyper parameters of the beta prior and of the Dirichlet prior, they should be compatible: hence the beta prior should be the one obtained from the Dirichlet by marginalisation.
\section{Conclusion}\label{crem}
This paper is intended to have several take-home messages. 
The first one is that a forensic statistician before starting any evaluation should make up his mind if he wants to use frequentist or Bayesian methods, since we have seen that the corresponding likelihood ratios are differently defined.
If a Bayesian approach is chosen, which has the advantage that everything is combined into a single number, without any uncertainty involved, the LR should be calculated in a principled way. 
Bayesian plug-in (and frequentist plug-in), often proposed as proper Bayesian approach, can sometimes be seen as a convenient approximation of the Bayesian LR, but this paper has shown that the full Bayesian method is not more difficult.  Moreover, the Bayesian plug-in is almost always anti-conservative in a way that is unfair to defence, and there are sometimes significant differences with the full Bayesian method for particular choices of the hyper-parameters of the priors.
All this has been shown when the conventional choices for the priori (beta or Dirichlet) are made. 
The choice of the prior is an issue indeed. We believe that a true Bayesian should not make use of conventional priors, but of his own priors. Especially because, as shown, this conventional choice leads to likelihood ratios which strongly depend on the hyperparameters of these priors. 
Choosing more realistic prior may increase the difficulty of the computation of the likelihood ratio, but, also thanks to modern computational tools, this should not stop people from preferring them.
%
%
%In the future, the use of asymmetric Dirichlet distribution will be investigated, along with alternative ways to build the prior distribution such as the Chinese Restaurant process \citep{pitman:2006}, and the Bayesian solution of \citet{brenner:2010}.

%Lastly, it is important to point out that in the Beta binomial model the prior is chosen after the evidence is observed, which is kind of contradicting of the notion of a prior. }
%For each method, the obtained LR values are compared to classical the plug-in estimates proposed in literature. The difference is not significant for the Beta binomial model, while for the Dirichlet - multinomial, can attain to almost two orders of magnitudes, and it can be seen that the plug-in estimate is always anti conservative if compared to the true LR. 

\section*{Acknowledgements}

I am indebted to Charles Brenner for the useful discussion about this paper, which lead to many improvements.
This research was supported by the Swiss National Science Foundation, through grants no.\ 105311-1445570 and 10531A-156146/1, and carried out in the context of a joint research project, supervised by Franco Taroni (University of Lausanne, Ecole des sciences criminelles), and Richard Gill (Mathematical Institute, Leiden University).

\end{document}